\documentclass[prc,twocolumn,showpacs,preprintnumbers,amsmath,amssymb,superscriptaddress,floatfix]{revtex4-2}

\usepackage{graphicx}
\usepackage{graphics}
\usepackage{subfigure}
\usepackage{dcolumn}
\usepackage{bm}
\usepackage[dvipsnames]{xcolor}
\colorlet{RED}{red}
\usepackage[T1]{fontenc}
\usepackage[utf8]{inputenc}
\usepackage{hyperref}
\hypersetup{
 colorlinks,
 linkcolor={red},
 citecolor={blue},
 urlcolor={blue}
}
\usepackage{tabularx}

\usepackage{float}
\newcommand{\brfive}{\texorpdfstring{$^{85}$Br }{85Br }}
\newcommand{\brseven}{\texorpdfstring{$^{87}$Br }{87Br }}
\newcommand{\brsevenc}{\texorpdfstring{$^{87}$Br}{87Br}}
\newcommand{\brnine}{\texorpdfstring{$^{89}$Br }{89Br }}
\newcommand{\brninec}{\texorpdfstring{$^{89}$Br}{89Br}}
\newcommand{\brnineone}{\texorpdfstring{$^{91}$Br }{91Br }}
\newcommand{\brnineonec}{\texorpdfstring{$^{91}$Br}{91Br}}

\newcommand{\brninethree}{\texorpdfstring{$^{93}$Br }{93Br }}
\newcommand{\brninethreec}{\texorpdfstring{$^{93}$Br}{93Br}}

\newcommand{\gr}{$\gamma$-ray }

\newcolumntype{Y}{>{\centering\arraybackslash}X}

\begin{document}

\preprint{APS/123-QED}

\title{High-resolution spectroscopy of neutron-rich Br isotopes and signatures for a prolate-to-oblate shape transition at N=56}

\newcommand{\IPHC}{Universit\'e de Strasbourg, CNRS, IPHC UMR 7178, F-67000 Strasbourg, France}
\newcommand{\IPTwoI}{Universite Claude Bernard Lyon 1, CNRS/IN2P3, IP2I Lyon, UMR 5822, Villeurbanne, F-69100, France}
\newcommand{\GANIL}{GANIL, CEA/DRF-CNRS/IN2P3, BP 55027, 14076 Caen cedex 5, France}
\newcommand{\CSNSM}{CSNSM, CNRS/IN2P3, Universit\'e Paris-Saclay, 91405 Orsay, France}
\newcommand{\TUDarmstadt}{Institut f\"ur Kernphysik, Technische Universit\"at Darmstadt, D-64289 Darmstadt, Germany}
\newcommand{\IFIC}{Instituto de F\'isica Corpuscular, CSIC-Universitat de Val\`encia, E-46980 Valencia, Spain}
\newcommand{\Canada}{Department of Chemistry, Simon Fraser University, Burnaby, British Columbia, Canada}
\newcommand{\Legnaro}{INFN, Laboratori Nazionali di Legnaro, I-35020 Legnaro, Italy}
\newcommand{\IPN}{Institut de Physique Nucl\'eaire, CNRS/IN2P3, Universit\'e Paris Saclay, 91406 Orsay Cedex, France}
\newcommand{\Debrecen}{HUN-REN Institute for Nuclear Research (HUN-REN ATOMKI), Pf.51, H-4001, Debrecen, Hungary}
\newcommand{\Somerset}{iThemba LABS, National Research Foundation, P.O.Box 722, Somerset West,7129 South Africa}
\newcommand{\Padova}{INFN Sezione di Padova, I-35131 Padova, Italy}
\newcommand{\UPadova}{Dipartimento di Fisica e Astronomia dell'Universit\`a di Padova, I-35131 Padova, Italy}
\newcommand{\GSI}{GSI, Helmholtzzentrum f\"ur Schwerionenforschung GmbH, D-64291 Darmstadt, Germany}
\newcommand{\Milano}{INFN, Sezione di Milano, I-20133 Milano, Italy}
\newcommand{\LPSC}{LPSC, Universit\'e Grenoble-Alpes, CNRS/IN2P3, 38026 Grenoble Cedex, France}
\newcommand{\IRFU}{IRFU, CEA/DRF, Centre CEA de Saclay, F-91191 Gif-sur-Yvette Cedex, France}
\newcommand{\UMilano}{Dipartimento di Fisica, Università degli Studi di Milano, 20133 Milano, Italy}
\newcommand{\STFC}{STFC Daresbury Laboratory, Daresbury, Warrington, WA4 4AD, UK}
\newcommand{\ILL}{Institut Laue-Langevin, Grenoble, France}
\newcommand{\IFINHH}{Horia Hulubei National Institute of Physics and Nuclear Engineering—IFIN HH, Bucharest 077125, Romania}
\newcommand{\IJCLAB}{Université Paris-Saclay, CNRS/IN2P3, IJCLab, 91405 Orsay, France}


\author{J.~Dudouet}
\email[Corresponding author: ]{j.dudouet@ip2i.in2p3.fr}
\affiliation{\IPTwoI}

\author{G. Colombi}
\altaffiliation[Present Address: ]{Dept. of Physics, University of Guelph, 50 Stone Road East, Guelph, ON N1G2W1 Canada}
\affiliation{\ILL}
\affiliation{\UMilano}

\author{D. Reygadas Tello}
\author{C. Michelagnoli}
\affiliation{\ILL}

\author{D.D.~Dao} 
\author{F.~Nowacki} 
\affiliation{\IPHC}


\author{M. Abushawish} 
\affiliation{\IPTwoI}

\author{E. Clément} 
\affiliation{\GANIL}

\author{C. Costache} 
\affiliation{\IFINHH}

\author{G. Duchêne} 
\affiliation{\IPHC}

\author{F. Kandzia} 
\affiliation{\ILL}


\author{A. Lemasson} 
\affiliation{\GANIL}

\author{N. Marginean}
\author{R. Marginean}
\author{C. Mihai}
\author{S. Pascu}
\affiliation{\IFINHH}

\author{M. Rejmund} 
\affiliation{\GANIL}

\author{K. Rezynkina} 
\altaffiliation[Present Address: ]{\Padova}
\affiliation{\IPHC}%

\author{O. Stezowski} 
\affiliation{\IPTwoI}%

\author{A. Turturica} 
\author{S. Ujeniuc}
\affiliation{\IFINHH}



\author{A.~Astier}
\affiliation{\IJCLAB}
\affiliation{\CSNSM}

\author{G.~de~Angelis}
\affiliation{\Legnaro}

\author{G.~de~France}
\affiliation{\GANIL}

\author{C.~Delafosse}
\affiliation{\IJCLAB}
\affiliation{\IPN} 

\author{I.~Deloncle}
\thanks{CSNSM and IPNO no longer exist, they have been replaced by IJCLab in 2020}
\affiliation{\IJCLAB} 
\affiliation{\CSNSM}


\author{A.~Gadea}
\affiliation{\IFIC}

\author{A.~Gottardo}
\altaffiliation[Present Address: ]{\Legnaro}
\thanks{CSNSM and IPNO no longer exist, they have been replaced by IJCLab in 2020}
\affiliation{\IPN} 


\author{P.~Jones}
\affiliation{\Somerset}

\author{T.~Konstantinopoulos}
\thanks{CSNSM and IPNO no longer exist, they have been replaced by IJCLab in 2020}
\affiliation{\CSNSM}

\author{I.~Kuti}
\affiliation{\Debrecen}

\author{F.~Le~Blanc}
\affiliation{\IPHC}
\affiliation{\IJCLAB}

\author{S.M.~Lenzi}
\affiliation{\Padova}
\affiliation{\UPadova}


\author{R.~Lozeva}
\affiliation{\IPHC}
\affiliation{\IJCLAB} 
\affiliation{\CSNSM}

\author{B.~Million}
\affiliation{\Milano}



\author{R.M.~P\'erez-Vidal}
\altaffiliation[Present Address: ]{\Legnaro}
\affiliation{\IFIC}

\author{C.M.~Petrache}
\affiliation{\IJCLAB} 
\affiliation{\CSNSM}

\author{D.~Ralet}
\thanks{CSNSM and IPNO no longer exist, they have been replaced by IJCLab in 2020}
\affiliation{\TUDarmstadt}
\affiliation{\CSNSM} 


\author{N. Redon}  
\affiliation{\IPTwoI}%

\author{C.~Schmitt}
\affiliation{\IPHC}

\author{D.~Sohler}
\affiliation{\Debrecen}

\author{D.~Verney}
\affiliation{\IJCLAB}
\affiliation{\IPN}

\begin{abstract} 
The first systematic experimental study of the neutron-rich Br isotopes with two  complementary state-of-the-art techniques is presented. These isotopes have been populated in the fission process at two different facilities, GANIL and ILL. New spectroscopic information has been obtained for odd-even $^{87-93}$Br isotopes and the experimental results have been compared with state-of-the-art Large-Scale Shell-Model  and DNO Shell-Model calculations. As a result of such theoretical approaches, a transition from prolate ($^{87,89}$Br) to oblate ($^{91,93}$Br) shapes is obtained from the subtle balance between proton and neutron quadrupole deformations, as a clear signature of pseudo-SU3 quadrupole regime. 
\end{abstract}

\maketitle

\section{Introduction}\label{sec:introduction}

Excited states in odd mass nuclei, such as the odd-even Br isotopes provide an ideal testing ground for single-particle energies and the coupling of single particle states with core vibrations. With 35 protons, the Br isotopes are ideal probes for the proton-neutron interaction between the $\pi fp$ and $\nu dg$ orbitals -- a key ingredient for modern Shell Model calculations in the mass region~\cite{SORLIN2008602}. Such calculations can be found in the literature in different experimental works. High-spin excitations of \brfive\ ($N=50$) have been studied using fusion-fission reactions in~\cite{Astier_2006,Fotiades_2005}. Both the ground state and first excited states were associated with protons in the $\pi f_{5/2}$, $\pi p_{3/2}$, $\pi p_{1/2}$, and $\pi g_{9/2}$ orbitals. More recently, \brseven ($N=52$) has been extensively studied from $\beta$-decay~\cite{wisniewski_2019} and $^{235}$U neutron-induced fission~\cite{Nyako_2021}. Excited states associated with single proton excitations were identified. A good agreement between measured and predicted excitation energies was found for negative-parity states. In contrast, positive-parity states (built on the $\pi g_{9/2}$ configuration) could not be reproduced. The odd-odd $^{86,88}$Br isotopes were also studied using fusion fission reactions~\cite{Porquet2009} and, more recently, neutron-induced fission~\cite{Jentschel_2017} and theoretically investigated with Large-Scale Shell Model (LSSM) calculations~\cite{Czerwinski_2015,Urban_2016}. Finally, \brnine ($N=54$) has been very recently studied for the first time, here also from $\beta$-decay~\cite{Urban_2022} and fission~\cite{Nyako_2021_2}. It has been shown that \brseven and \brnine  have similar level schemes, while a smooth enhancement of the collectivity seems to appear, based on Shell Model interpretation. An isomeric nature of the 9/2$^{+}$ state was recently suggested~\cite{wisniewski_2019}, in order to explain the transition strengths observed in the $\beta$-decay experiment. A lifetime of 20\,ns was suggested, implying a well defined structure for the low-lying states in such isotopes. 

From the measurement of atomic masses of neutron-rich $^{85–92}$Br and $^{94–97}$Rb isotopes a reduction of the shell gap at $Z=35$ compared to $Z=40$ was observed as well as a poor agreement of the theoretical shell gap values with the experimental results~\cite{Rahaman07}. 

In the recent Monte Carlo Shell Model~\cite{OTSUKA2001319} or DNO-LSSM~\cite{Dao_2022} calculations, the collective effects are taken into account. Mean-field methods are used to calculate potential energy surfaces and ``preselect'' a basis for SM calculations. Such approach recently revealed successful to describe a transition from semi-magicity to $\gamma$-softness in the $^{83-87}$As nuclei~\cite{Rezynkina_2022}. In particular, the new spectroscopic data obtained in that work were interpreted in terms of pseudo-SU3 symmetries, pointing to the moderate prolate deformation for the ground states of $^{85}$As and $^{87}$As. A description of $^{85}$As and $^{87}$As as deformed nuclei was given using DNO-SM calculations. The excited states in $^{85}$As have been reported to have a prolate-triaxial nature, while $^{87}$As (just one proton below $^{86}$Ge, expected to have maximum triaxial deformation~\cite{Rezynkina_2022}) has been predicted to be $\gamma$ soft.

With an additional proton, similar effects of evolution of deformation are expected for the Br isotopes. A prolate to oblate transition has been predicted at $N=58$ in neutron-rich Kr isotopes, on the basis of Symmetry Conserving Configuration Mixing (SCCM) calculations with the Gogny D1S interaction~\cite{Rodriguez2014}.
It is worth noting that in recent studies neutron-rich $^{88-94}$Se isotopes both triaxial degree of freedom and shape coexistence were found to play an important role in the description of intrinsic deformations in such isotopes~\cite{Chen_2017}. 

\section{\label{sec:setup}Experimental setup}

In this work, excited states in neutron-rich odd Br isotopes were populated via fission reactions and studied using two complementary \gr spectroscopy experiments.

A first measurement was performed at GANIL~\footnote{Data from the E680 GANIL experiment} using the combination of the large-acceptance magnetic spectrometer VAMOS++~\cite{Rejmund_2011} and the AGATA $\gamma$-ray tracking array~\cite{Akkoyun_2012,Clement_2017}. A $^{238}$U beam, accelerated up to a kinetic energy of 6.2 MeV/u and with an average intensity of $\sim 1$~pnA, was used to impinged on a 10 $\mu$m thick $^{9}$Be target. VAMOS++ was used to obtain an event-by-event determination of the mass (A) and atomic number (Z) of the detected fragments~\cite{Navin_2014}. Prompt $\gamma$ rays emitted at the target position were detected in coincidence with AGATA, composed of 8 triple clusters (24 HPGe crystals), placed in compact configuration (13.3~cm from the target). The positions of the $\gamma$-ray interaction points were determined via pulse-shape analysis techniques~\cite{Bruyneel_2016,Venturelli_2004}. The $\gamma$-ray path in the Ge array was reconstructed using a tracking algorithm~\cite{Lopez-Martens_2004} to obtain the total $\gamma$-ray energy along with the position of its first interaction. The combination of the measurement of the velocity vector of the fission fragments using VAMOS++ and the determination of the position of the first interaction of the $\gamma$ ray in AGATA allowed to apply a precise event-by-event Doppler correction. As a result, despite a measured $v/c$ of $\approx0.1$, a $\gamma$-ray energy resolution of 5~keV (FWHM) has been obtained at 1.2 MeV. Further details of the experimental setup and characterisation can be found in \cite{Rejmund_2011,Navin_2014,Lemasson2023,Kim2017}. This data-set was used to identify transitions in Br isotopes with no known spectroscopic information, as well as to extend the level scheme of known isotopes. 

\begin{figure}[ht!]
\includegraphics[width=\linewidth]{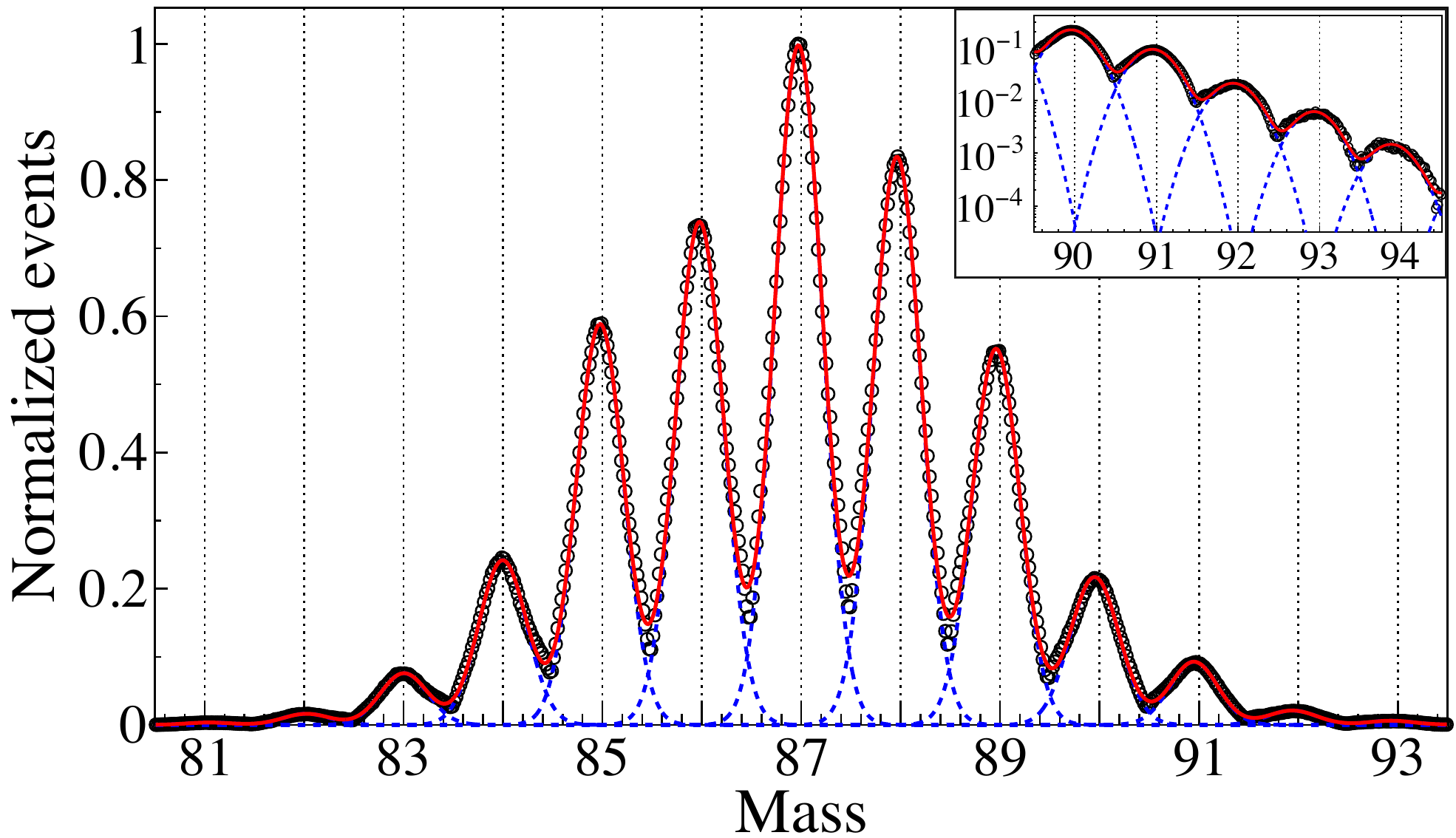}%
\caption{(color online) Normalized mass distribution for $Z=35$ isotopes detected in the VAMOS++ spectrometer. The continuous red and dotted blue lines represent the fit of the total distributions and individual mass contributions, respectively.}
\label{fig:Br_ADist}
\end{figure}

Fig.~\ref{fig:Br_ADist} shows the atomic mass distribution for $Z=35$ isotopes that have been observed in coincidence with $\gamma$ rays. The figure illustrates the quality of the isotopic identification and the yields of the different isotopes identified in the spectrometer. A systematic check of the analysis procedure and of the isotopic selectivity has been performed on the data, already used for extracting spectroscopic results on the $^{96}$Kr, $^{81}$Ga and As nuclei~\cite{Dudouet_2017,Dudouet_2019,Rezynkina_2022}.

\begin{figure*}
\includegraphics[width=\linewidth]{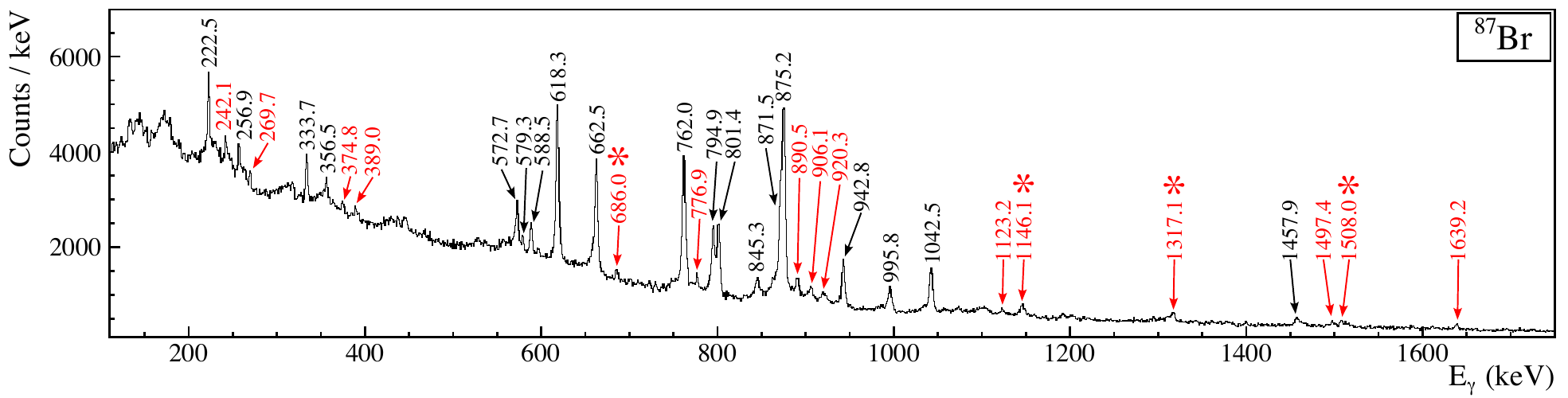}%
\caption{(color online) Tracked $\gamma$-ray spectrum of $^{87}$Br obtained  in the  AGATA-VAMOS++ experiment. Newly observed transitions are labeled in red, and an asterisk is added for the transitions placed in the level scheme. All the other transitions are already known. There are no transitions from contaminants.}
\label{fig:87Br_AV_spectrum}
\end{figure*}

Another data-set was obtained from a neutron-induced fission experimental campaign performed with FIPPS (FIssion Product Prompt $\gamma$-ray Spectrometer) at ILL~\cite{Michelagnoli_2018, DOI_FissionExpFIPPS}. A collimated thermal neutron beam of $\sim 10^8$  neutrons cm$^{-2}$s$^{-1}$, was used to induce fission on a $^{235}$U target. For the first time at a neutron beam facility, an active fission target~\cite{Kandzia_2020} was used. This allowed a selection of the fission events and a suppression of $\beta$-decay induced $\gamma$-ray background. Prompt $\gamma$ rays from the fission fragments were detected using a HPGe array consisting of 16 HPGe clovers installed around the target position (8 HPGe clovers from the FIPPS instrument complemented by additional 8 HPGe clover detectors equipped with anti-Compton shields, from the IFIN-HH collaboration). The data were recorded using a digital acquisition system, in trigger-less mode. An event-builder was run offline, using the active target signal as a trigger and requiring a $\gamma$ multiplicity of at least two (in a time window of 400\,ns). The data were sorted offline into $\gamma-\gamma-\gamma$ cubes (1x10$^{11}$ events). The $\gamma$ coincidence technique was employed to identify new transitions and place them in the level schemes. Data were analysed after applying the add-back procedure to $\gamma$ rays detected in the crystals belonging to the same clover within a time interval of 250\,ns. An overall energy resolution of 2.4~keV (FWHM) at 1.4~MeV was achieved for the FIPPS data.

The \gr energies reported in this work are taken from the AGATA data. The experimental errors on those quantities result from the combination of the statistical error, obtained from the fit of the peaks in the energy spectrum, together with the systematic error. The latter has been obtained from the comparison of the literature and experimental values for approximately 50 among the most intense transitions observed in the same data-set.  
A systematic error of 0.2~keV has been deduced from this analysis (mainly due to kinematics reconstruction uncertainties). The excitation energies and their error have been obtained via an average taking into account the different $\gamma$-ray decay paths.

The presented relative intensity values for the different transitions are corrected for the detector efficiencies, and the uncertainties include the statistical error and systematic one (of 5\%, coming from the detector efficiency correction based on GEANT4 simulations). 

\section{Experimental results}\label{sec:results}

\subsection{$^{87}_{35}$Br$_{52}$}\label{subsec:brseven}

\begin{figure}[htb]
\includegraphics[width=\linewidth]{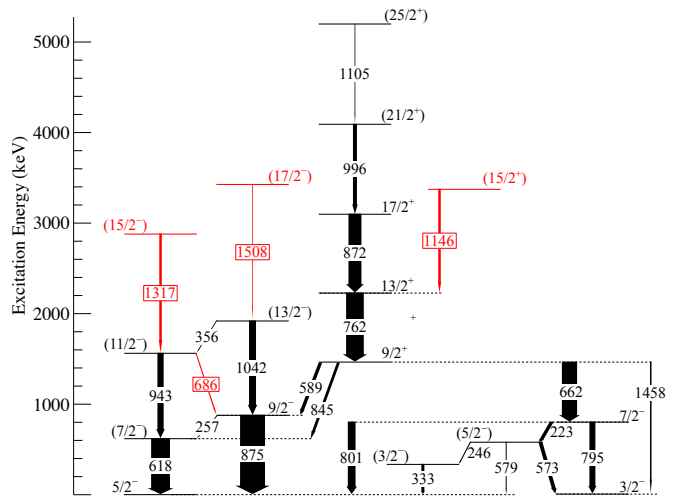}%
\caption{(color online) Level scheme for \brseven obtained in this work. New transitions and levels are shown in red boxes. The width of the arrows reflects the observed intensities relative to the strongest transition. The intensities were obtained using the AGATA singles spectrum.}
\label{fig:87Br_LS}
\end{figure}

Situated just two neutrons above the $N=50$ shell closure, the \brseven nucleus was recently studied following fission reactions during the EXILL campaign~\cite{Jentschel_2017,Nyako_2021}. Excited states populated in the $\beta$-decay of $^{87}$Se were also extensively studied in~\cite{wisniewski_2019}. From the combined analysis of the AGATA-VAMOS++ and FIPPS data-sets, the known level scheme of \brseven was confirmed and extended in this work (see Fig.~\ref{fig:87Br_LS}).

\begin{figure}[htb]
\includegraphics[width=\linewidth]{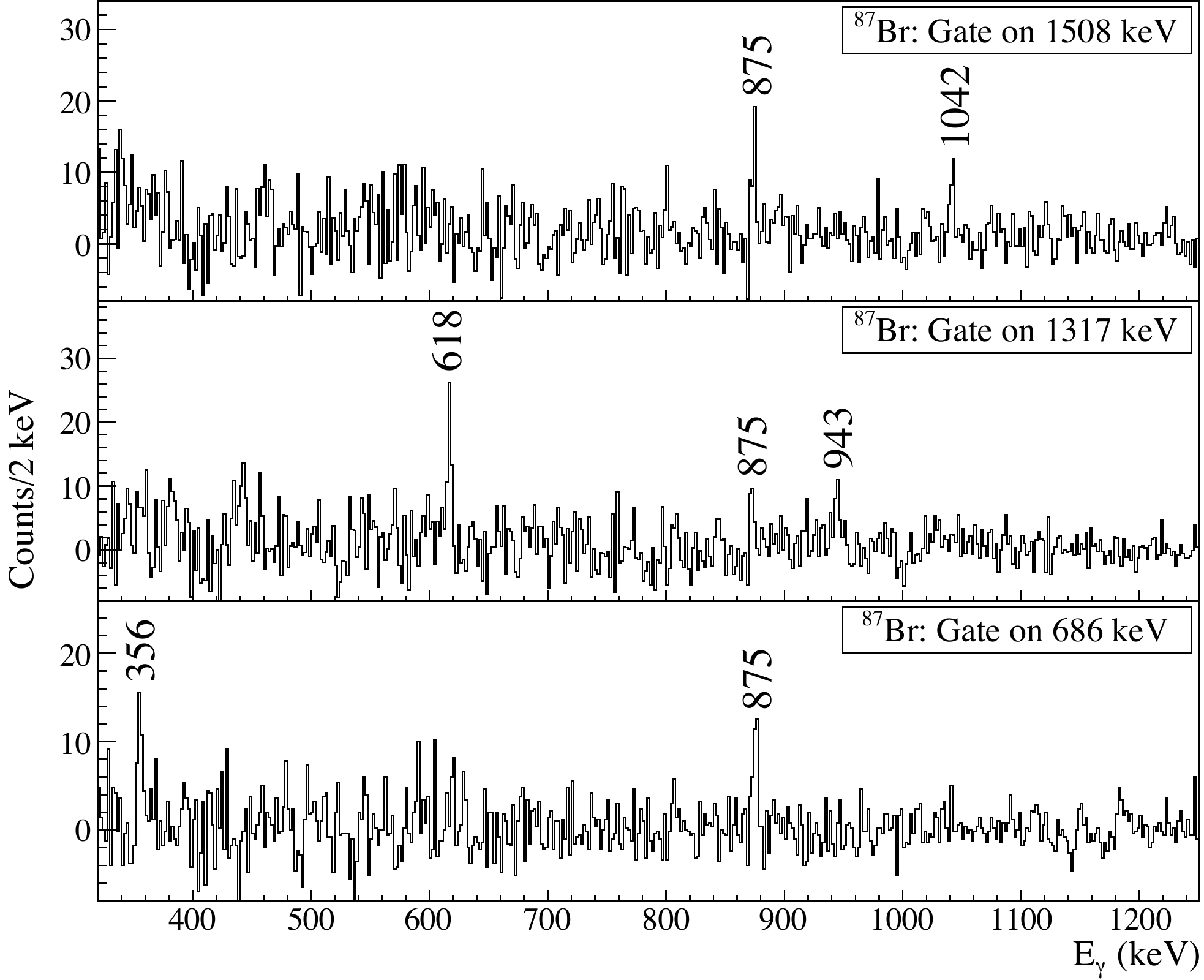}%
\caption{Example of tracked $\gamma$-$\gamma$ coincidence spectra obtained in the  AGATA-VAMOS++ experiment confirming the placement of newly observed transitions in the \brseven level scheme.}
\label{fig:87Br_Gates}
\end{figure}

The $\gamma$-ray spectrum obtained for \brseven with the AGATA-VAMOS++ setup is shown in Fig.~\ref{fig:87Br_AV_spectrum}, on which all the newly observed transitions are labeled. Asterisks indicate transitions that have been added to the level scheme in this work. The non-labeled peaks correspond to already known $\gamma$ rays. Among the 15 new transitions attributed to this nucleus, only 4 were populated with sufficient statistics to allow for coincidences and, thus, be placed in the level scheme. The updated level scheme is shown in Fig.~\ref{fig:87Br_LS} where the new transitions are presented in red boxes. An example of coincidence spectra, obtained with AGATA, confirming the placement of the 1317 and 1508~keV transitions, is shown in Fig.~\ref{fig:87Br_Gates}.

\begin{figure}[htb]
\includegraphics[width=\linewidth]{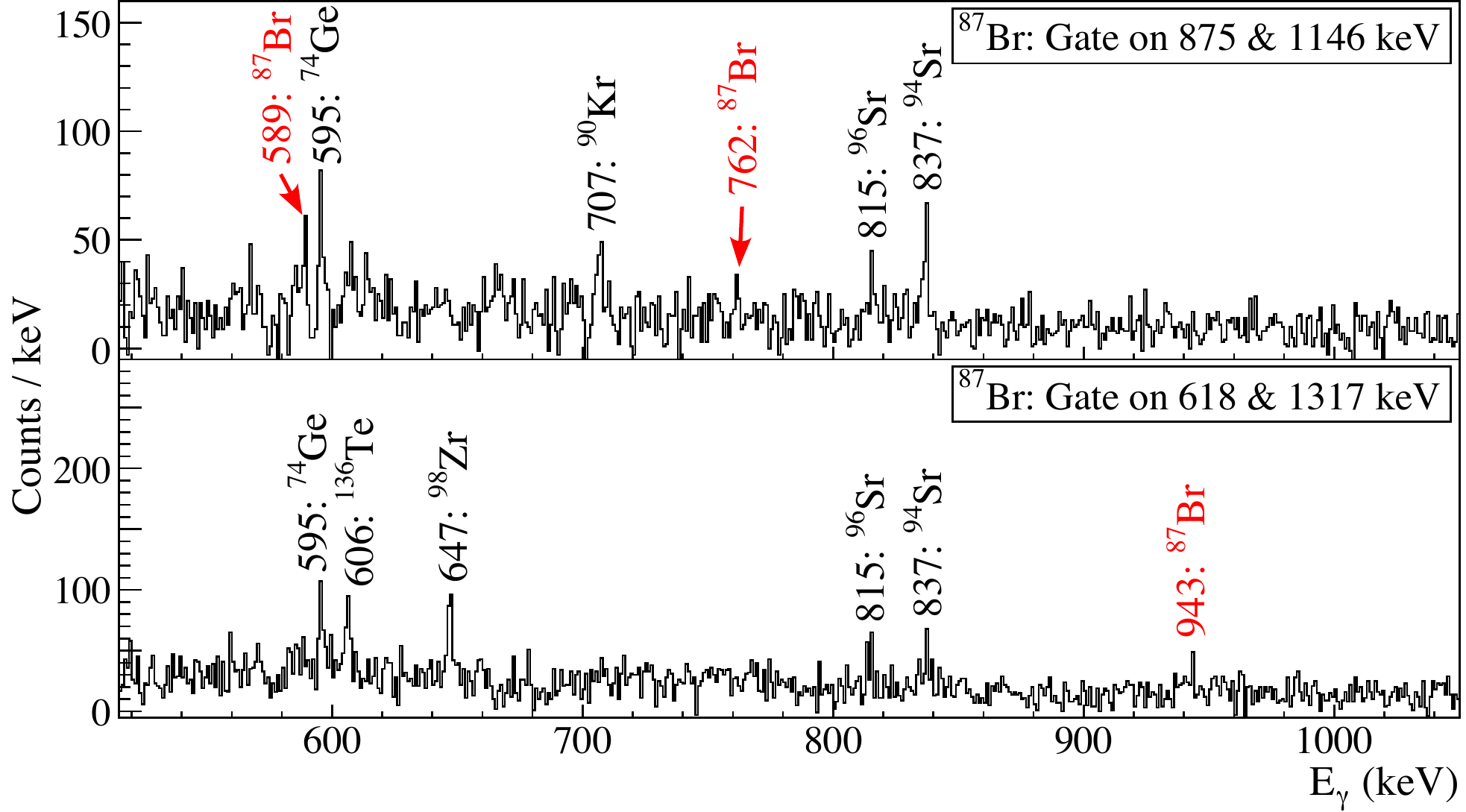}%
\caption{(color online) Example of $\gamma$-$\gamma$-$\gamma$ coincidence spectra obtained with FIPPS confirming the placement of newly observed transitions in the \brseven level scheme.}
\label{fig:87Br_Gates_FIPPS}
\end{figure}

In the case of \brsevenc, the FIPPS data-set was able to confirm the recently published level scheme from~\cite{Nyako_2021}. Among the newly observed transitions with AGATA, this data-set allowed to place the 1146~keV transition in the level scheme and to confirm the placement of the 1317~keV one, as shown in Fig.~\ref{fig:87Br_Gates_FIPPS}. The different contaminants labeled in Fig.~\ref{fig:87Br_Gates_FIPPS} come from different sources. $^{73}$Ge(n,$\gamma$) reactions on the detectors produce a $\gamma$ ray of 595.5~keV that can be seen as random coincidences. $^{98}$Zr and its partner $^{136}$Te, $^{90}$Kr and $^{94-96}$Sr are among the most produced isotopes, contributing to a small fraction of the overall background.

The spin assignment for the new $(15/2^-)$ and $(17/2^-)$ states was made assuming a dominant population of the yrast states in fission reactions, resulting in a ``continuation'' of the two E2 yrast negative parity bands. The comparison to the Large Scale Shell Model (LSSM) calculations (more details in the following, see Fig.~\ref{fig:SM_calcs_comparison}) fully supports these tentative assignments. A spin $(15/2^+)$ was also tentatively assigned to a new level, at an energy of 3372~keV, on the basis of LSSM calculations. The $(15/2^+_1)$ state is indeed predicted to be at an energy of 3213~keV, above the $(17/2^+_1)$ predicted at 3124~keV (compared to 3097~keV for the experimental value).

The derived level energies for \brsevenc, $\gamma$-ray energies, relative intensities and their placement in the level scheme are summarized in Tab.~\ref{tab:87Br}.

\begin{table}[ht!]
\begin{tabularx}{\linewidth}{XYXXX}
\hline\hline
$E_i$ (keV) & $J_i^\pi$     & $E_\gamma$ (keV)          & $E_f$ (keV)       & $I_\gamma$        \\
\hline                              
0           & $5/2^-_1$     & -                         &  -                &     -             \\
6.2(6)      & $3/2^-_1$     & -                         &  -                &     -             \\
333.6(3)    & $(3/2^-_2)$   & 333.6(3)$^{a}$            &  0                &   12(1)           \\
579.3(7)    & $(5/2^-_2)$   & 572.7(3)                  &  6.2              &   16(2)           \\
            &               & 579.3(3)                  &  0                &   3.7(4)          \\
            &               & 246(1)$^{a}$              &  333.6            &   <1              \\
618.3(2)    & $(7/2^-_1)$   & 618.3(2)                  &  0                &   74(6)           \\
801.6(4)    & $(7/2^-_2)$   & 801.4(3)                  &  0                &   31(3)           \\
            &               & 794.8(3)                  &  6.2              &   24(2)           \\
            &               & 222.5(3)                  &  579.3            &   16(2)           \\
875.2(3)    & $(9/2^-_1)$   & 875.2(3)                  &  0                &   100              \\
            &               & 256.9(3)                  &  618.3            &   5.0(5)          \\
1463.8(4)   & $(9/2^+_1)$   & 662.4(2)                  &  801.6            &   61(5)           \\
            &               & 588.5(3)                  &  875.2            &   14(2)           \\
            &               & 845.3(4)                  &  618.3            &   12(1)           \\
            &               & 1457.9(5)$^{a}$           &  6.2              &   6.1(7)          \\
1561.2(5)   & $(11/2^-_1)$  & 942.8(3)                  &  618.3            &   25(2)           \\
            &               & \textbf{686.0(5)}$^{a}$   &  \textbf{875.2}   &   \textbf{5.2(6)} \\
1917.6(4)   & $(13/2^-_1)$  & 1042.4(3)$^{a}$           &  875.2            &   28(3)           \\
            &               & 356.5(3)$^{a}$            &  1561.2           &   3.8(4)          \\
2225.8(5)   & $(13/2^+_1)$  & 762.0(2)                  &  1463.8           &   71(6)           \\
2878.2(8)   & $(15/2^-_1)$  & \textbf{1317.1(6)}        &  \textbf{1561.2}  &   \textbf{9.5(9)} \\
3097.3(8)   & $(17/2^+_1)$  & 871.5(6)                  &  2225.8           &   52(4)           \\
3371.9(6)   & $(15/2^+_1)$  & \textbf{1146.1(4)}        &  \textbf{2225.8}  &   \textbf{9.7(9)} \\
3425.6(1.1) & $(17/2^-_1)$  & \textbf{1508(1)}$^{a}$    &  \textbf{1917.6}  &   \textbf{2.7(4)} \\
4093.1(8)   & $(21/2^+_1)$  & 995.8(3)                  &  3097.3           &   16(2)           \\
5198.1(8)   & $(25/2^+_1)$  & 1105.0(2)$^{f}$           &  4093.1           &   -   \vspace{5pt}\\
\multicolumn{5}{l}{Unplaced transitions} \\
 -          & -             &  242.1(3)$^{a}$           &  -                &   4.4(5)          \\ 
 -          & -             &  269.7(3)$^{a}$           &  -                &   4.2(4)          \\ 
 -          & -             &  374.8(5)$^{a}$           &  -                &   3.2(4)          \\ 
 -          & -             &  389.0(4)$^{a}$           &  -                &   1.7(3)          \\ 
 -          & -             &  776.9(3)$^{a}$           &  -                &   3.8(4)          \\ 
 -          & -             &  890.5(3)$^{a}$           &  -                &   11(1)           \\ 
 -          & -             &  906.1(5)$^{a}$           &  -                &   8.7(8)          \\ 
 -          & -             &  920.3(6)$^{a}$           &  -                &   8.4(8)          \\ 
 -          & -             &  1123.2(6)$^{a}$          &  -                &   3.1(4)          \\ 
 -          & -             &  1497.4(4)$^{a}$          &  -                &   2.7(4)          \\ 
 -          & -             &  1639.2(5)$^{a}$          &  -                &   4.0(5)          \\ 
\hline\hline
\end{tabularx}
\caption{Level energies, spin-parities, $\gamma$-ray energies and relative intensities corresponding to the observed level scheme of \brsevenc. Intensities are given relatively to the strongest transition intensity from the AGATA singles spectrum. A label is added to transitions that have been observed only with AGATA$^{a}$ or only with FIPPS$^{f}$. The new transitions placed in the level scheme are reported in bold.}
\label{tab:87Br}
\end{table}

\subsection{$^{89}_{35}$Br$_{54}$}\label{subsec:brnine} 

\begin{figure*}[ht!]
\includegraphics[width=\linewidth]{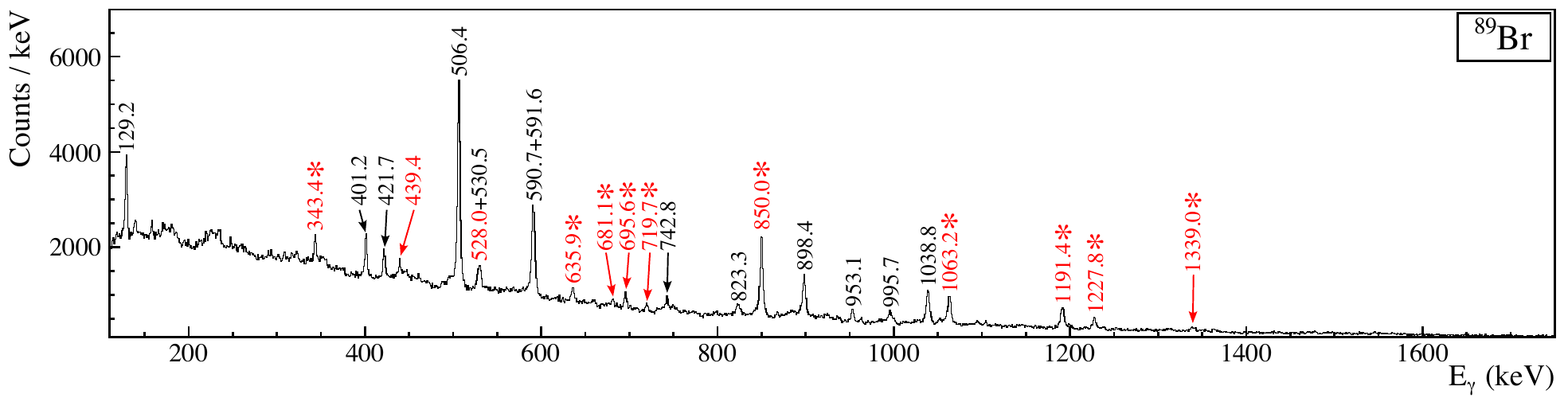}%
\caption{(color online) Tracked $\gamma$-ray spectrum of $^{89}$Br obtained  in the  AGATA-VAMOS++ experiment. Newly observed transitions are labeled in red, and an asterisk is added for the transitions placed in the level scheme. All the other transitions are already known. There are no transitions from contaminants.}
\label{fig:89Br_AV_spectrum}
\end{figure*}

\begin{figure}[htb]
\includegraphics[width=\linewidth]{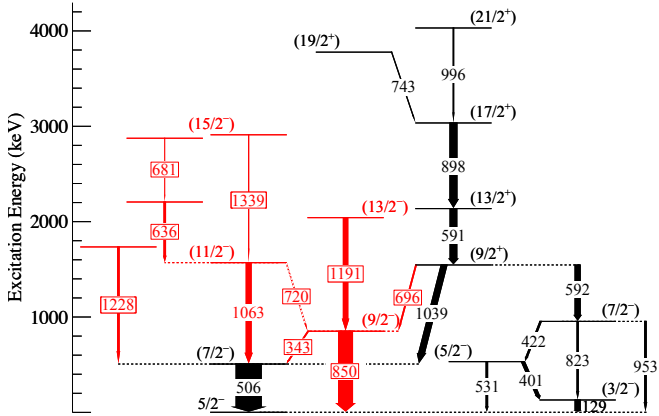}%
\caption{(color online) Level scheme for \brnine obtained in this work. New transitions and levels are shown in red boxes. The width of the arrows reflects the observed intensities relatively to the strongest transition. The intensities were obtained using the AGATA singles spectrum. See text for details on the spin and parity assignments. The 720~keV transition is shown in dashed line because the statistics didn't allow to observe it in coincidences. The placement is only based on the energy difference of the excited states.}
\label{fig:89Br_LS}
\end{figure}

Before performing the two experiments considered in this work, no spectroscopic information was known about $^{89}$Br $(N=54)$. First results on the spectroscopy of \brnine have been recently published in \cite{Nyako_2021_2}. These results have been confirmed and extended in this work (see Fig.~\ref{fig:89Br_LS}). The \gr spectrum obtained for \brnine with the AGATA-VAMOS++ setup is shown in Fig.~\ref{fig:89Br_AV_spectrum}, on which all the newly observed transitions are labeled. Asterisks have been added for those transitions that have been added to the level scheme in this work. Among the 12 newly discovered transitions attributed to this nucleus, 10 could be placed in the level scheme. The updated level scheme is shown in Fig.~\ref{fig:89Br_LS}, where the new transitions are highlighted with red boxes.

\begin{figure}[htb]
\includegraphics[width=\linewidth]{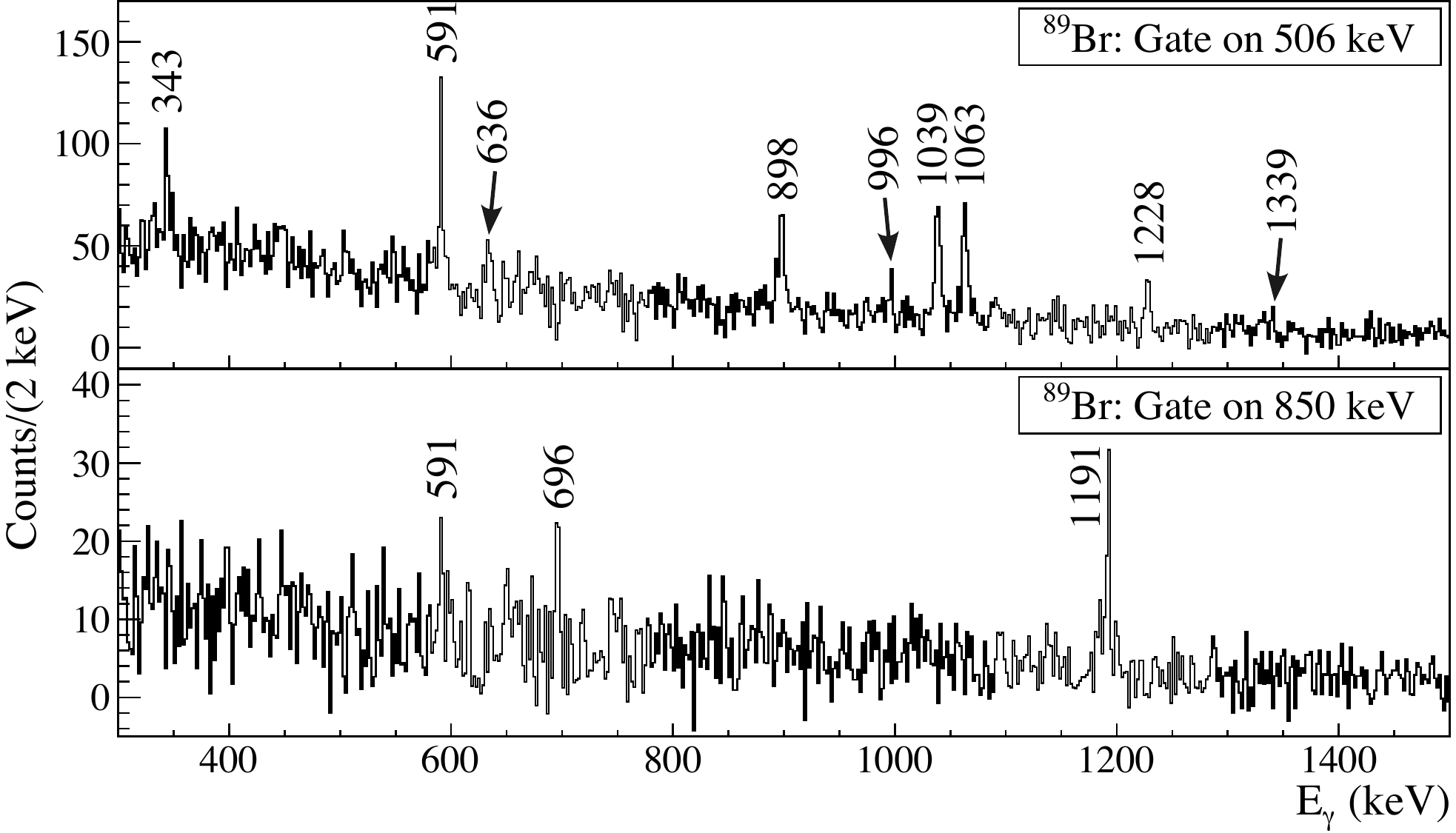}%
\caption{Example of tracked $\gamma$-$\gamma$ coincidence spectra obtained in the  AGATA-VAMOS++ experiment confirming the placement of newly observed transitions in the \brnine level scheme.}
\label{fig:89Br_Gates}
\end{figure}

\begin{figure}[htb]
\includegraphics[width=\linewidth]{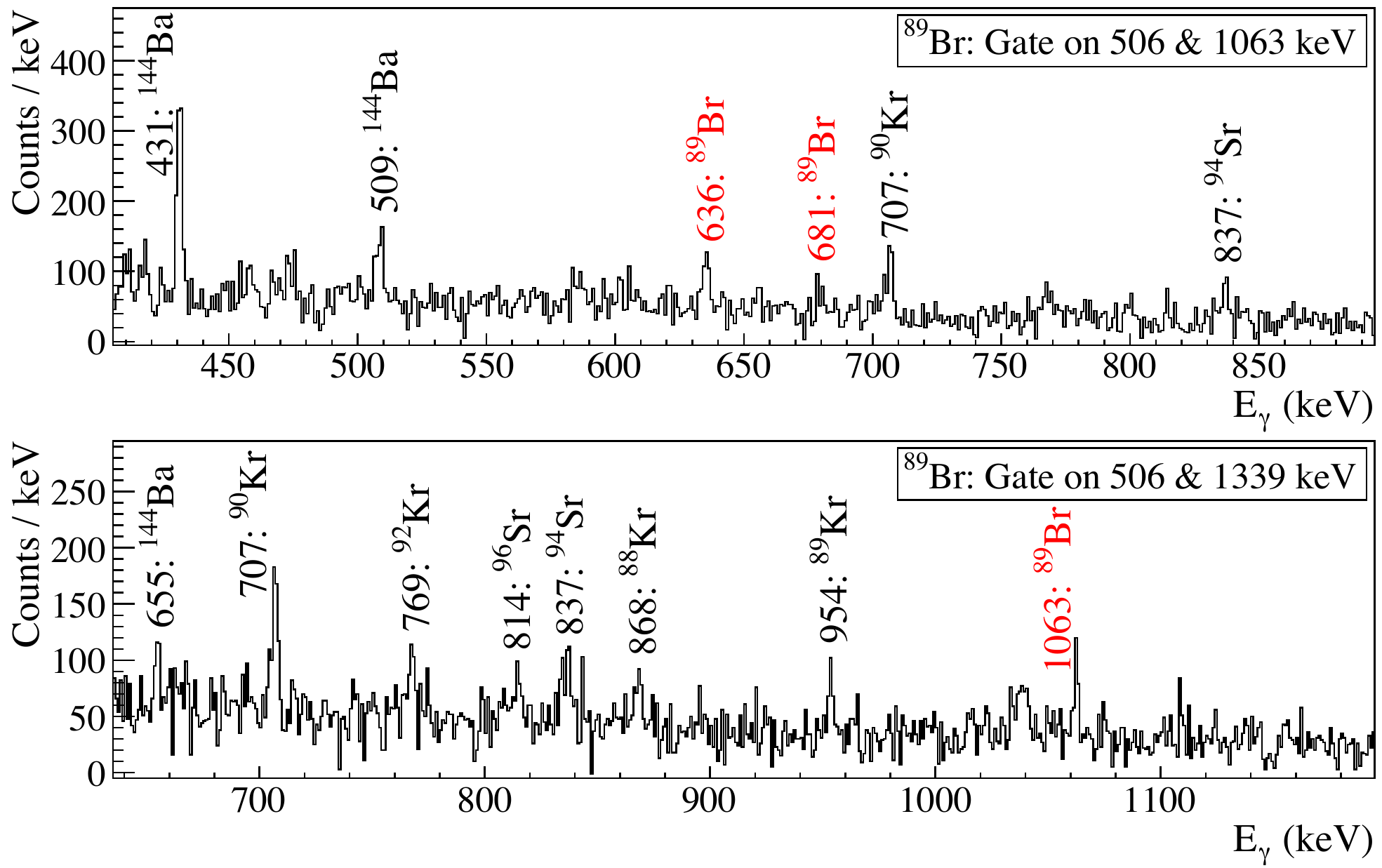}%
\caption{(color online) Example of $\gamma$-$\gamma$-$\gamma$ coincidence spectrum obtained with FIPPS confirming the placement of newly observed transitions in the \brnine level scheme.}
\label{fig:89Br_FIPPS_Gates}
\end{figure}

The level scheme has been built based on the combination of $\gamma-\gamma$ coincidences using AGATA and $\gamma$-$\gamma$-$\gamma$ coincidences using FIPPS. Examples of coincidence spectra obtained with AGATA and FIPPS are shown in Fig.~\ref{fig:89Br_Gates} and Fig.~\ref{fig:89Br_FIPPS_Gates}, respectively. $^{144}$Ba is one of the dominant fission fragments and it is characterized by a transition at 509~keV which explains its important contribution to the gated spectra of Fig.~\ref{fig:89Br_FIPPS_Gates}, as random coincidence, as well as its most probable complementary fragments: $^{88,89,90}$Kr~\cite{Schmidt_2016}. $^{94,96}$Sr are also among the most produced isotopes, contributing to a small fraction of the overall background. Finally, $^{92}$Kr is not produced with a dominant yield, but its level scheme contains a transition at 506~keV, explaining the presence of the 769~keV transition that is the most intense transition emitted by this nucleus. While the placement of the 1339~keV transition is not convincing from the AGATA analysis (Fig.~\ref{fig:89Br_Gates}, top panel), the double-gate spectrum obtained with FIPPS (Fig.~\ref{fig:89Br_FIPPS_Gates}, bottom panel) allows for a firm placement.


One of the transitions placed in the level scheme, of 720~keV energy, with very low statistics, has not been seen in coincidence but it has been placed tentatively based on the energy difference of the excited states and on the basis of the systematics, given that a similar transition is observed in \brsevenc. 

In the two published studies on \brninec, the first one, from the EXILL campaign~\cite{Jentschel_2017} was performed from neutron-induced fission reaction, without isotopic identification of the emitting fragment and with no information on the \brnine level scheme~\cite{Nyako_2021_2}. The use of the $\gamma$ rays emitted by the fission partners  allowed to find ``pairs'' of possible candidates for decays of \brnine. Due to these experimental limitations, the decay from the $(9/2^-)$ state at 850~keV could not be observed, despite its strong intensity in fission. The second study was a $\beta^-$-decay experiment \cite{wisniewski_2019}. With a ground state of spin-parity of $3/2^+$, the $\beta^-$-decay of $^{89}$Se to the $(9/2^-)$ state in \brnine was not observed. Thanks to the isotopic identification obtained with VAMOS++, this strong transition at 850~keV was measured in this work for the first time, and it has been attributed to the $(9/2^-_1)$ state according to the systematics in energy and intensity compared to \brseven. Two E2 yrast negative bands based on the $(7/2^-_1)$ and $(9/2^-_1)$ levels have been identified, in analogy to \brseven. This implies the new tentative assignments $(11/2^-)$, $(13/2^-)$ and $(15/2^-)$. The comparison to the Large Scale Shell Model (LSSM) calculations (see Fig.~\ref{fig:SM_calcs_comparison}) fully supports these tentative assignments. No spin was assigned for levels built on the 1228, 636, and 681~keV transitions. No clear evidence from systematics or from the LSSM calculations has been found to provide a tentative assignment.

The derived level energies for \brninec, $\gamma$-ray energies, relative intensities and their placement in the level scheme are summarized in Tab.~\ref{tab:89Br}.

\begin{table}[ht]
\begin{tabularx}{\linewidth}{XYXXX}
\hline\hline
$E_i$ (keV) & $J_i^\pi$          & $E_\gamma$ (keV)         & $E_f$ (keV)       & $I_\gamma$            \\
\hline         
0                & $5/2^-_1$     & -                        &  -                &     -                 \\
129.2(3)         & $(3/2^-_1)$   & 129.2(3)$^{a}$           &  0                &     24(2)             \\
506.4(2)         & $(7/2^-_1)$   & 506.4(2)                 &  0                &     100               \\
530.5(5)         & $(5/2^-_2)$   & 401.2(3)$^{a}$           &  129.2            &     17(2)             \\
                 &               & 530.5(6)$^{a}$           &  0                &     10(3)             \\
849.9(3)         & $(9/2^-_1)$   & \textbf{850.0(2)}$^{a}$  &  \textbf{0}       &     \textbf{57(5)}    \\
                 &               & \textbf{343.4(3)}$^{a}$  &  \textbf{506.4}   &     \textbf{8.3(8)}   \\
952.6(5)         & $(7/2^-_2)$   & 953.1(3)$^{a}$           &  0                &     11(1)             \\
                 &               & 421.7(3)$^{a}$           &  530.5            &     9.9(9)            \\
                 &               & 823.3(4)$^{a}$           &  129.2            &     8.9(9)            \\
1545.0(7)        & $(9/2^+_1)$   & 1038.8(3)                &  506.4            &     28(3)             \\
                 &               & 591.6(9)$^{a}$           &  952.6            &     25(2)             \\
                 &               & \textbf{695.6(3)}$^{a}$  &  \textbf{849.9}   &     \textbf{7.3(7)}   \\
1569.6(4)        & $(11/2^-_1)$  & \textbf{1063.2(3)}       &  \textbf{506.4}   &     \textbf{23(2)}    \\
                 &               & \textbf{719.7(4)}$^{a}$  &  \textbf{849.9}   &     \textbf{4.8(6)}   \\
1734.2(4)        & -             & \textbf{1227.8(3)}       &  \textbf{506.4}   &     \textbf{10(1)}    \\
2041.3(4)        & $(13/2^-_1)$  & \textbf{1191.4(3)}$^{a}$ &  \textbf{849.9}   &     \textbf{21(2)}    \\
2135.7(9)        & $(13/2^+_1)$  & 590.7(6)                 &  1545.0           &     30(3)             \\
2205.5(5)        & -             & \textbf{635.9(3)}        &  \textbf{1569.6}  &     \textbf{9.6(9)}   \\
2886.6(8)        & -             & \textbf{681.1(6)}        &  \textbf{2205.5}  &     \textbf{3.2(4)}   \\
2908.6(1.1)      & $(15/2^-_1)$  & \textbf{1339(1)}         &  \textbf{1569.6}  &     \textbf{4.1(5)}   \\
3034.1(0.9)      & $(17/2^+_1)$  & 898.4(3)                 &  2135.7           &     32(3)             \\
3776.9(1.0)      & $(19/2^+_1)$  & 742.8(3)$^{a}$           &  3034.1           &     4.6(5)            \\
4029.8(1.0)      & $(21/2^+_1)$  & 995.7(3)                 &  3034.1           &     6.1(7)\vspace{5pt}\\
\multicolumn{5}{l}{Unplaced transitions} \\
 -               & -             & 439.4(3)$^{a}$           &  -                &     3.7(4)            \\ 
 -               & -             & 528(2)$^{a}$             &  -                &     7(2)              \\          
\hline\hline
\end{tabularx}
\caption{Level energies, spin-parities, $\gamma$-ray energies and relative intensities corresponding to the observed level scheme of \brninec. Intensities are given relatively to the strongest transition intensity from the AGATA singles spectrum. A label is added to transitions that have been observed only with AGATA$^{a}$. The new transitions placed in the level scheme are reported in bold.}
\label{tab:89Br}
\end{table}

\subsection{$^{91}_{35}$Br$_{56}$}\label{subsec:brnineone} 

\begin{figure*}[htb]
\includegraphics[width=\linewidth]{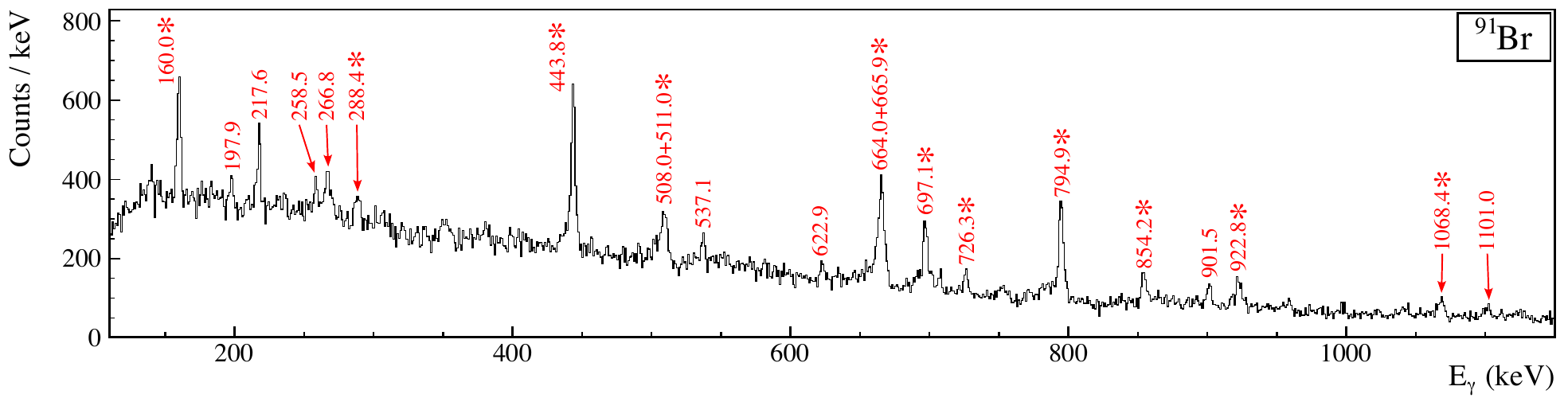}%
\caption{(color online) Tracked $\gamma$-ray spectrum of $^{91}$Br obtained  in the  AGATA-VAMOS++ experiment. Newly observed transitions are labeled, and an asterisk is added for the transitions placed in the level scheme.}
\label{fig:91Br_AV_spectrum}
\end{figure*}

\begin{figure}[htb]
\includegraphics[width=\linewidth]{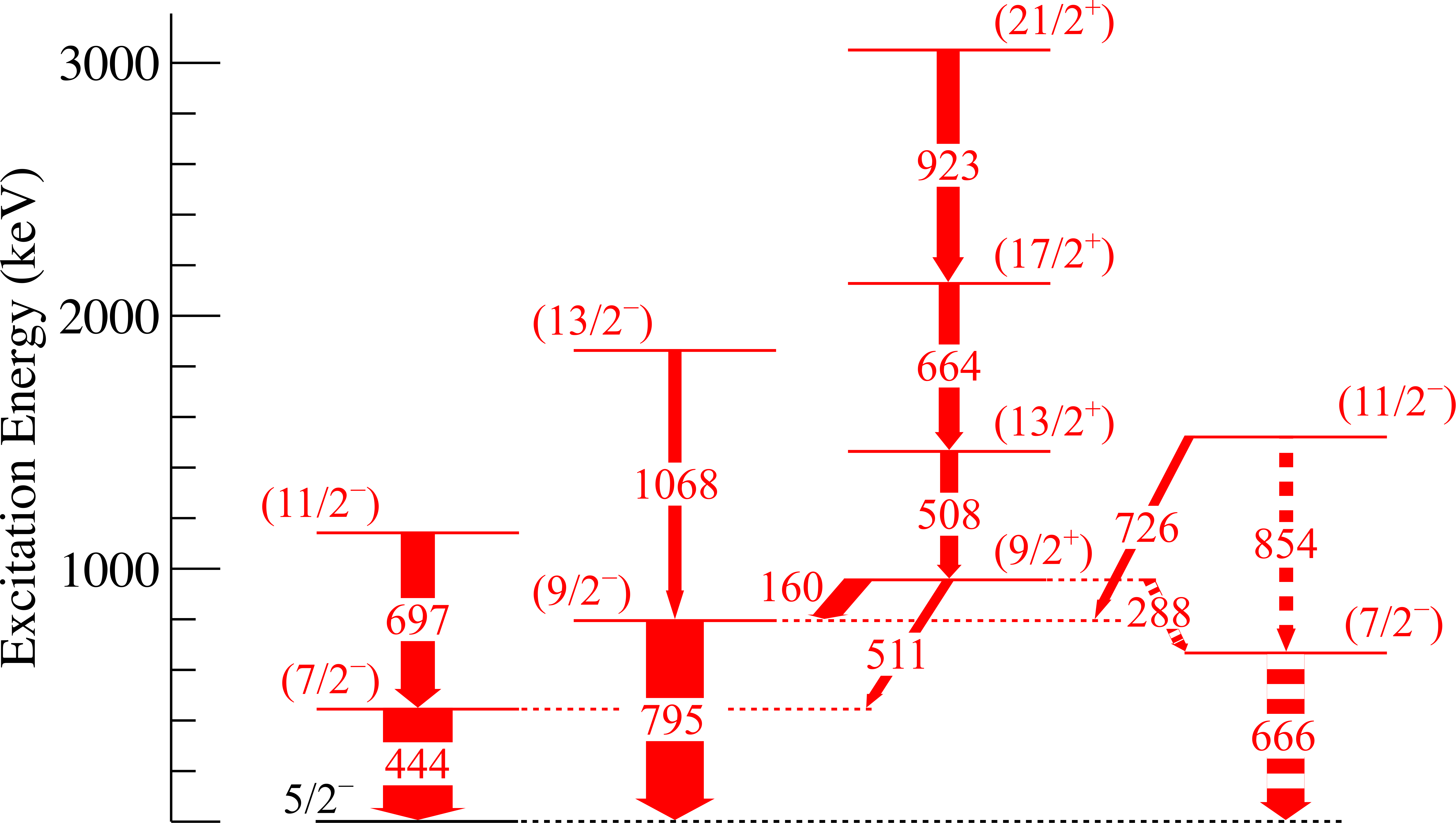}%
\caption{(color online) Level scheme for \brnineone obtained in this work for the first time. The width of the arrows reflects the observed intensities relative to the strongest transition. The intensities were obtained using the AGATA singles spectrum. See text for details on the spin and parity assignments. The 666~keV and 854~keV are seen in coincidence, but they are shown in dashed line because no coincidence was seen with the rest of the level scheme. The placement is only based on the energy matching, as for the 288~keV transition for which the statistics did not allow for coincidences.}
\label{fig:91Br_LS}
\end{figure}

No spectroscopic information was known for $^{91}$Br prior to this work. Due to its low production rate in the neutron-induced fission of $^{235}$U, an order of magnitude smaller than the $^{89}$Br, this isotope was not observed in the FIPPS data-set. Nevertheless, the transfer-fission mechanism through the $^{239}$U$^*$~compound~\cite{Ramos_2019}, together with the isotopic selectivity provided by VAMOS++ enabled an identification of the main $\gamma$-ray transitions in $^{91}$Br. The Doppler corrected prompt $\gamma$-ray spectrum measured in AGATA in coincidence with the $^{91}$Br ions detected in VAMOS++ is shown in Fig.~\ref{fig:91Br_AV_spectrum}. Among the 21 new transitions attributed to this nucleus, 13 (shown with asterisks) were placed in the level scheme. This level scheme, determined for the first time, is shown in Fig.~\ref{fig:91Br_LS}.

\begin{figure}[htb]
\includegraphics[width=\linewidth]{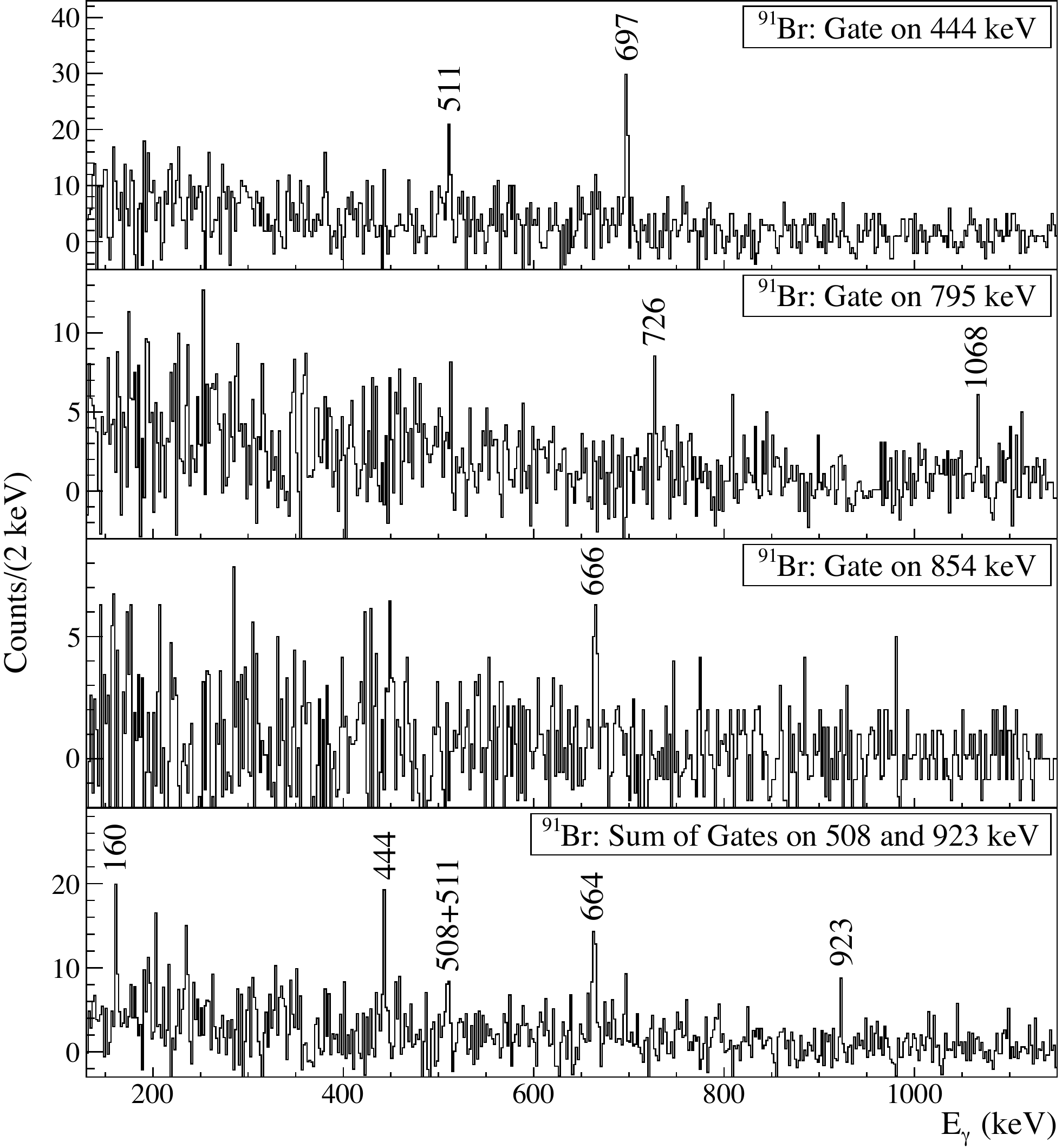}%
\caption{Example of tracked $\gamma$-$\gamma$ coincidence spectra obtained in the  AGATA-VAMOS++ experiment confirming the placement of newly observed transitions in the \brnineone level scheme.}
\label{fig:91Br_Gates}
\end{figure}

For this isotope, the coincidence analysis presented the challenge of limited statistics. As a result, some placements can only be tentative. The level scheme has been deduced from the combination of $\gamma-\gamma$ coincidences, energy matching and systematic studies (see section~\ref{sec:interpretation}). Some examples of coincidence spectra are shown in Fig.~\ref{fig:91Br_Gates}.

The clearest coincidences were observed between the 444 and 697~keV lines, while the two other main transitions, at 795 and 666~keV, were showing very low statistics in coincidence. While the 666 and 854~keV $\gamma$ rays were well seen in coincidence, the attribution of the 726 and 854~keV $\gamma$ rays to the same level at 1521~keV has only been made based on energy matching, as well as for the 288~keV $\gamma$ ray, for which the statistics was not sufficient to be observed in coincidence.

The construction and interpretation of the \brnineone level scheme has been particularly challenging. In addition to the low statistics obtained in $\gamma-\gamma$ coincidences, the presence of two doublets (508/511~keV) and (664/666~keV) has made the placement of the corresponding $\gamma$ rays in the level scheme quite complex. Moreover, the systematic comparison with the structures observed in \brseven and \brnine cannot be applied anymore. A structure similar to the one of the E2 yrast bands based on two main transitions de-exciting the $(7/2^-_1)$ and $(9/2^-_1)$ levels is still observed, but the structure based on the $(3/2^-_1)$, $(5/2^-_1)$ and $(7/2^-_2)$ levels in \brseven and \brnine are no more observed in \brnineonec, suggesting a deep modification of its nuclear structure. The systematic comparison with Se isotopes has been used to confirm the tentative spin-parity assignments of these three bands, based on the $(7/2^-)$, $(9/2^-)$ and $(9/2^+)$ levels (see section~\ref{sec:interpretation}). Considering this level scheme, and based on the fact that the $(9/2^+_1)$ state de-excitation mainly populates $(7/2^-)$ states in \brseven and \brninec, a tentative spin assignment of $(7/2^-_2)$ has been made for the level at 666~keV. Finally, based on the fact that the level at 1521~keV populates the $(7/2^-_2)$ and the $(9/2^-_1)$, and on the global systematic only reporting E2 and M1 transitions, its $(11/2^-_2)$ spin and parity assignment has been made tentatively. Despite these two last tentative attributions, it should be kept in mind that the position of the 666 and 1521~keV levels are not firmly established.

The derived level energies for \brnineonec, $\gamma$-ray energies, relative intensities and their placement in the level scheme are summarized in Tab.~\ref{tab:91Br}.

\begin{table}[ht]
\begin{tabularx}{\linewidth}{XYXXX}
\hline\hline
$E_i$ (keV) & $J_i^\pi$     & $E_\gamma$ (keV)   & $E_f$ (keV)   & $I_\gamma$   \\
\hline         
0           & $5/2^-_1$     & -                  &  -            &     -        \\
443.8(3)    & $(7/2^-_1)$   & 443.8(3)           &  0            &     100      \\
665.9(6)    & $(7/2^-_2)$   & 665.9(6)           &  0            &     54(8)    \\
794.9(3)    & $(9/2^-_1)$   & 794.9(3)           &  0            &     83(6)    \\
954.7(1.3)  & $(9/2^+_1)$   & 160.0(3)           &  794.9        &     44(4)    \\
            &               & 511(2)             &  443.8        &     16(5)    \\
            &               & 288.4(6)           &  665.9        &     18(2)    \\
1140.9(4)   & $(11/2^-_1)$  & 697.1(3)           &  443.8        &     49(4)    \\
1463(2)     & $(13/2^+_1)$  & 508(2)             &  954.7        &     26(6)    \\
1520.7(6)   & $(11/2^-_2)$  & 854.2(4)           &  665.9        &     21(2)    \\
            &               & 726.3(4)           &  794.9        &     15(2)    \\
1863.3(7)   & $(13/2^-_1)$  & 1068.4(6)          &  794.9        &     20(2)    \\
2127(3)     & $(17/2^+_1)$  & 664(2)             &  1463         &     30(6)    \\
3050(3)     & $(21/2^+_1)$  & 922.8(6)           &  2527         &     33(3)     \vspace{5pt}\\
\multicolumn{5}{l}{Unplaced transitions} \\
 -          & -             & 197.9(4)           &  -            &     12.8(9)  \\ 
 -          & -             & 217.6(3)           &  -            &     27(2)    \\ 
 -          & -             & 258.5(4)           &  -            &     8.0(6)   \\ 
 -          & -             & 266.8(6)           &  -            &     10.4(8)  \\ 
 -          & -             & 537.1(4)           &  -            &     14(1)    \\ 
 -          & -             & 622.9(4)           &  -            &     10.7(8)  \\ 
 -          & -             & 901.5(4)           &  -            &     17(2)    \\ 
 -          & -             & 1101(2)            &  -            &     13(1)    \\ 
\hline\hline
\end{tabularx}
\caption{Level energies, spin-parities, $\gamma$-ray energies and relative intensities corresponding to the observed level scheme of \brnineonec. Intensities are given relatively to the strongest transition intensity from the AGATA singles spectrum.}
\label{tab:91Br}
\end{table}

\subsection{$^{93}_{35}$Br$_{58}$}\label{subsec:brninethree} 

\begin{figure}[htb]
\includegraphics[width=\linewidth]{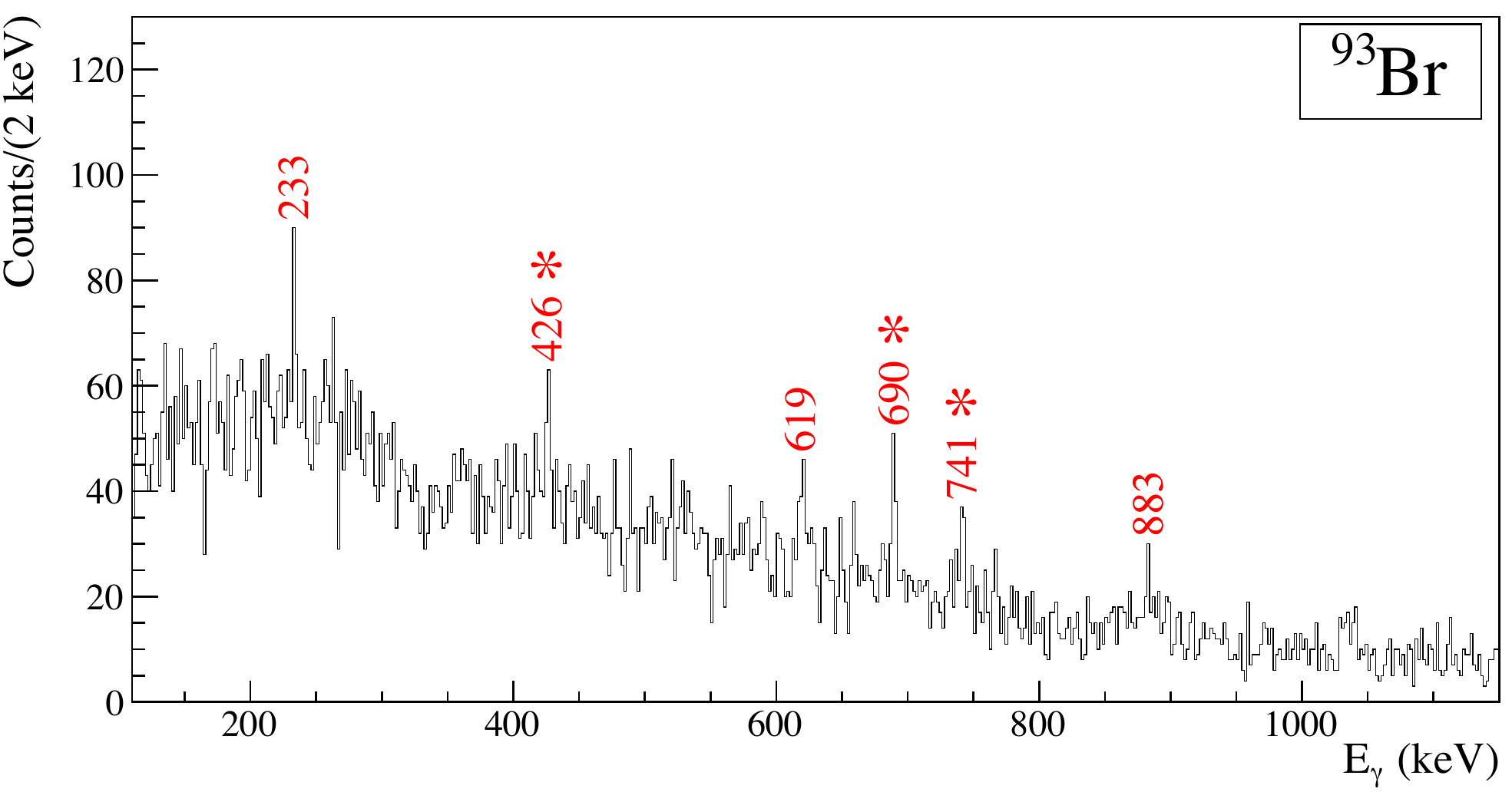}%
\caption{(color online) Tracked $\gamma$-ray spectrum of $^{93}$Br obtained in the AGATA-VAMOS++ experiment. Newly observed transitions are labeled and an asterisk is added for the transitions placed in the tentative level scheme.}
\label{fig:93Br_AV_spectrum}
\end{figure}

\begin{figure}[htb]
\includegraphics[width=0.6\linewidth]{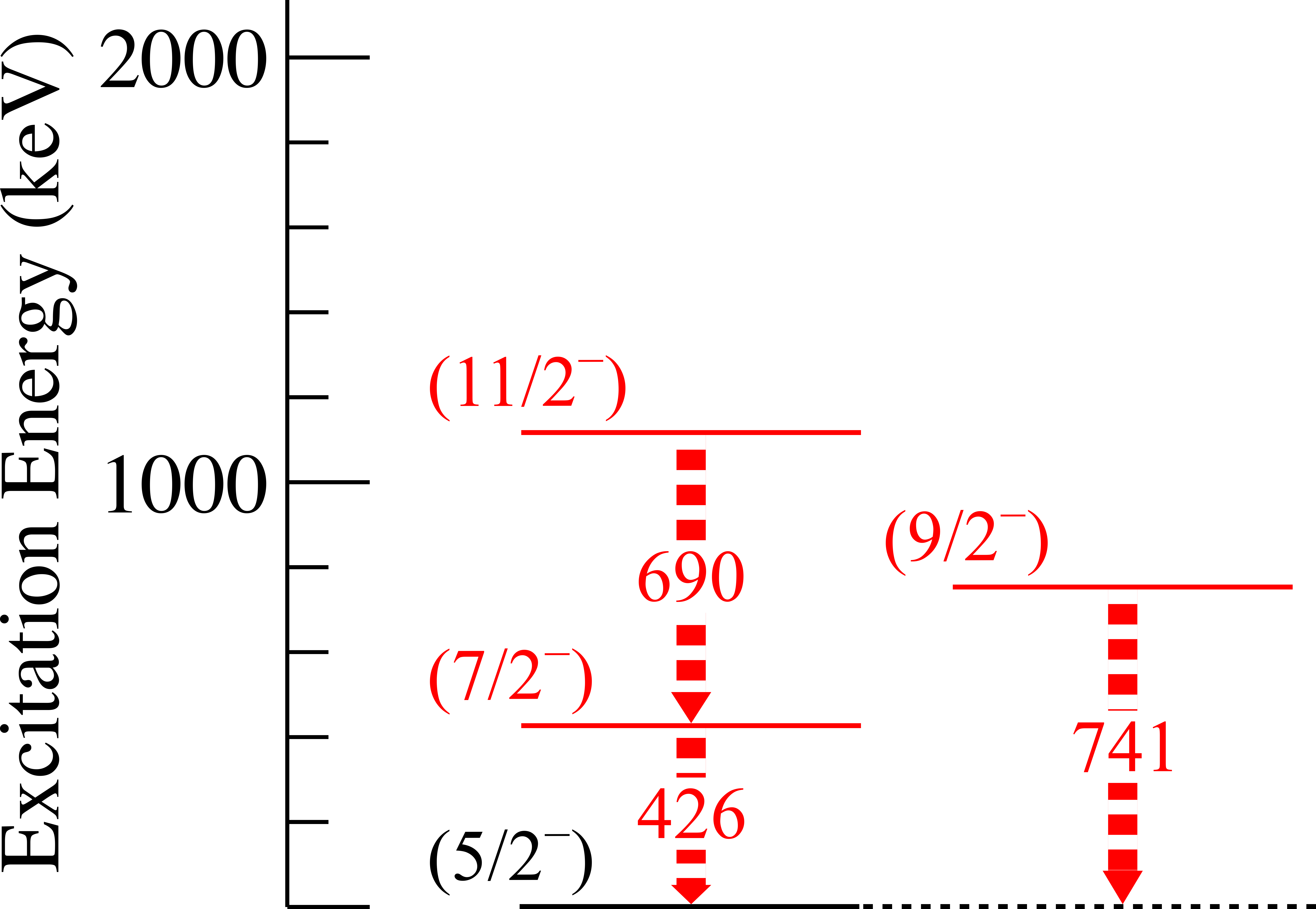}%
\caption{(color online) Tentative level scheme for \brninethree obtained in this work for the first time. Due to the impossibility of having coincidences with such low statistics, the proposed levels are only provisional and deduced from the systematic comparison with Se isotopes (see section \ref{sec:interpretation}). It was also not possible to extract relevant intensities. The width of the arrow is thus identical for all transitions. See text for details on the spin and parity assignments.}
\label{fig:93Br_LS}
\end{figure}

Located well above the $N=50$ shell closure, the \brninethree isotope is the most neutron-rich Br nucleus that could be studied in this work. Because of the same argument reported above for the \brnineone case, the FIPPS data-set could not be exploited. The Doppler corrected prompt $\gamma$-ray spectrum measured in AGATA in coincidence with the \brninethree ions detected in VAMOS++ is shown in Fig.~\ref{fig:93Br_AV_spectrum}. Due to the very low statistics obtained in the singles spectrum, a dedicated study has been made for this specific case to verify that none of the presented $\gamma$ rays could be originating from a contaminant. The produced spectrum has been made using very restrictive selections on the Z and A identifications, and it has been checked that none of these energies are present in the most likely contaminants ($^{92,93}$Kr from Z and mass contamination, $^{92}$Br from mass contamination and $^{89,90}$Br from charge state contamination). Among the 6 new transitions attributed to this nucleus, 3 (shown with asterisks in Fig.~\ref{fig:93Br_AV_spectrum}) were tentatively placed in the level scheme. The proposed provisional level scheme, measured for the first time, is shown in Fig.~\ref{fig:93Br_LS}. Due to the impossibility of carrying out a coincidence analysis because the too low statistics, the proposed spins and parity assignments are only tentative and deduced from the systematic comparison with Se isotopes (see Fig.~\ref{fig:system_Se} in section \ref{sec:interpretation}).

The derived level energies for \brninethreec, $\gamma$-ray energies and their tentative placement in the level scheme are summarized in Tab.~\ref{tab:93Br}.

\begin{table}[ht]
\begin{tabularx}{\linewidth}{XYXX}
\hline\hline
$E_i$ (keV) & $J_i^\pi$     & $E_\gamma$ (keV)   & $E_f$ (keV)    \\
\hline       
0           & $(5/2^-_1)$   & -                  &  -           \\
426(2)       & $(7/2^-_1)$   & 426(2)             &  0          \\
741(2)       & $(9/2^-_1)$   & 741(2)             &  0          \\
1116(3)      & $(11/2^-_1)$  & 690(2)             &  426         \vspace{5pt}\\
\multicolumn{4}{l}{Unplaced transitions} \\
 -          & -             & 233(2)             &  -           \\ 
 -          & -             & 619(2)             &  -          \\ 
 -          & -             & 883(2)             &  -           \\ 
\hline\hline
\end{tabularx}
\caption{Level energies, spin-parities, and $\gamma$-ray energies corresponding to the tentative level scheme of \brninethreec.}
\label{tab:93Br}
\end{table}


\subsection{Interpretation of the results in comparison with the systematics in the region}\label{sec:interpretation}

The excited states of the odd-even neutron-rich Br isotopes are distributed into three main bands, as evident from the previously shown level schemes (see Figs.~\ref{fig:87Br_LS}, \ref{fig:89Br_LS}, and \ref{fig:91Br_LS}). 

The band head of the first band is the $5/2^-$ ground state. This corresponds to an odd proton in the $\pi f_{5/2}$ orbital. As already discussed in~\cite{Nyako_2021_2}, excited states of $^{87}$Br are built on top of this level due to a coupling of the single particle configuration with vibrations of the $^{86}$Se core~\cite{Materna_2015}. The $\pi f_{5/2} \otimes (2^+,4^+,6^+)$ couplings give rise to the observed $7/2^-$ and $9/2^-$ states, $11/2^-$ and $13/2^-$ states, and $15/2^-$ and $17/2^-$ states, respectively. This interpretation, in terms of coupling with a Se core, could also be applied to the other Br isotopes. 

A second band is built on top of the first $3/2^-$ excited state, at 6~keV for $^{87}$Br and at 129~keV for $^{89}$Br. The coupling with the Se core, $\pi p_{3/2} \otimes 2^+$, gives rise to the observed $5/2^-_2$ and $7/2^-_2$ states. This band has not been observed for $^{91}$Br and $^{93}$Br, due either to the fact that, being non yrast, its population is not favored by fission processes, or to a change in the structure of the involved states.

A third band is built on top of the $9/2^+$ state and can be interpreted as a coupling with the Se core $\pi g_{9/2} \otimes 0^+$. Using the energies of the $2^+$, $4^+$ and $6^+$ states in the Se core, the levels above can be interpreted as $\pi g_{9/2} \otimes  (2^+,4^+,6^+)$, giving rise to $13/2^+$, $17/2^+$ and $21/2^+$ states. A similar interpretation has already been made for Rb~\cite{Fotiades_2005} and Y~\cite{Fotiades_2012} nuclei in this mass region. 

\begin{figure}[htb]
    \centering
    \includegraphics[width=\linewidth]{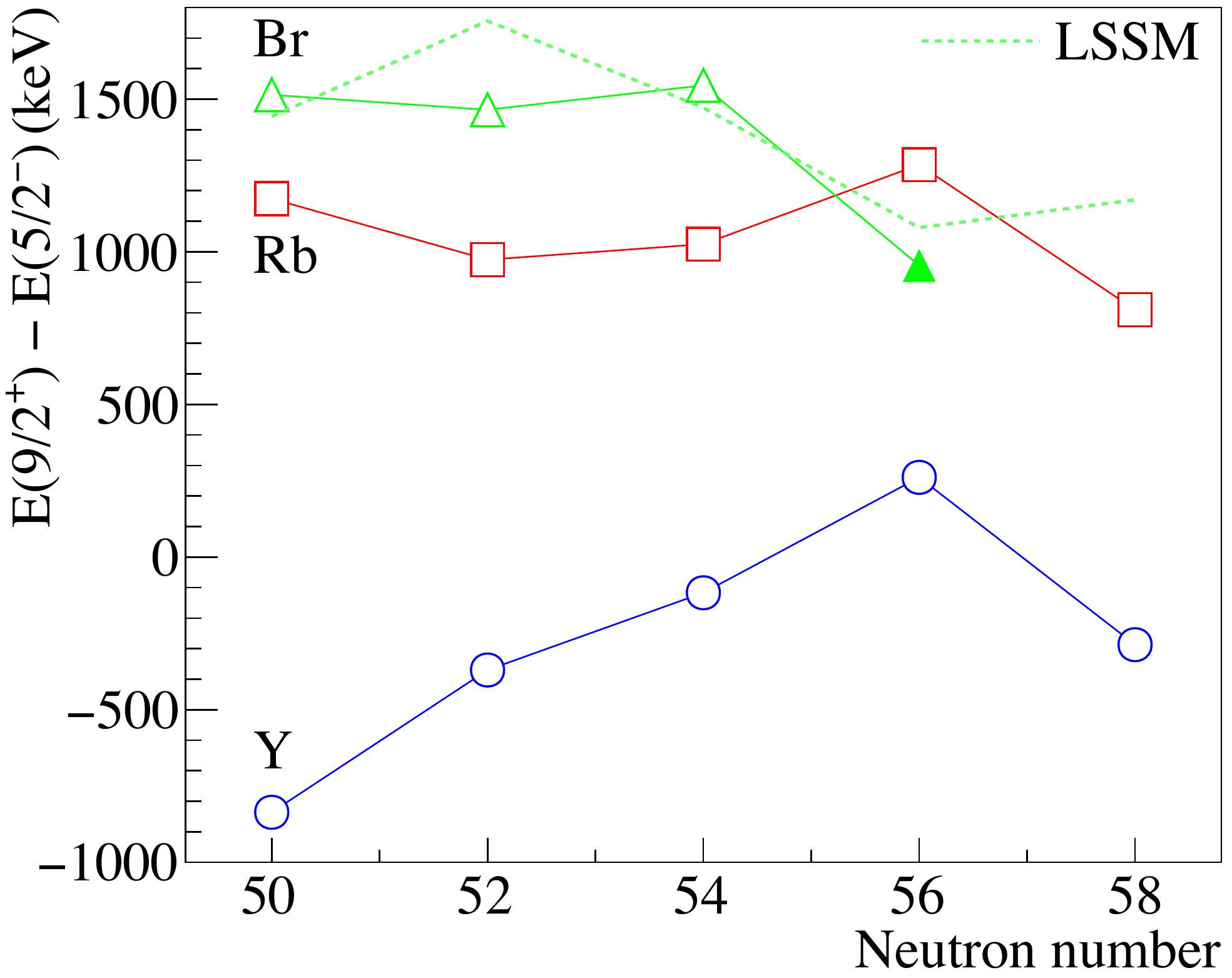}
   \caption{(color online) Systematics of the energy difference between the $9/2^+$ and $5/2^-$ states for the Br, Rb and Y isotopic chains. Empty symbols correspond to already published measurements, the filled symbol corresponds to the newly reported state in this work. Energy values are taken from~\cite{ensdf,Simpson2010}. The LSSM predictions from this work are represented with a dashed line.}
    \label{fig:systemBr_9_2_p}
\end{figure}

The evolution of the $9/2^+$ state deserves particular attention. The systematics of the difference between the energy of this state and the one of the first $5/2^-$ (not always the ground state in the considered nuclei) is shown in Fig.~\ref{fig:systemBr_9_2_p} for the Br ($Z=35$), Rb ($Z=37$) and Y ($Z=39$) isotopes. Empty symbols are used for the values extracted from the literature~\cite{ensdf,Simpson2010}, while filled symbol is used for the new result in \brnineonec. The trend shows a maximum at $N=56$ for the Y and Rb isotopes, which could be related to the $\nu d_{5/2}$ sub-shell closure. On the contrary, a ``drop'' in this systematics is observed for the Br isotopes at $N=56$, suggesting an important modification in the structure of those nuclei as compared to the higher-Z ones. This new finding adds evidence for a significant change in nuclear structure between $Z=35$ and $Z=37$, as demonstrated for the Kr isotopic chain ($Z=36$), nowadays known as the low-Z boundary of the $N=60$ island of deformation~\cite{Dudouet_2017}. The dashed line in Fig.~\ref{fig:systemBr_9_2_p} indicates a very good agreement with the Shell Model predictions from this work, discussed in section~\ref{sec:theory}.

\begin{figure}[htb]
    \centering
    \includegraphics[width=\linewidth]{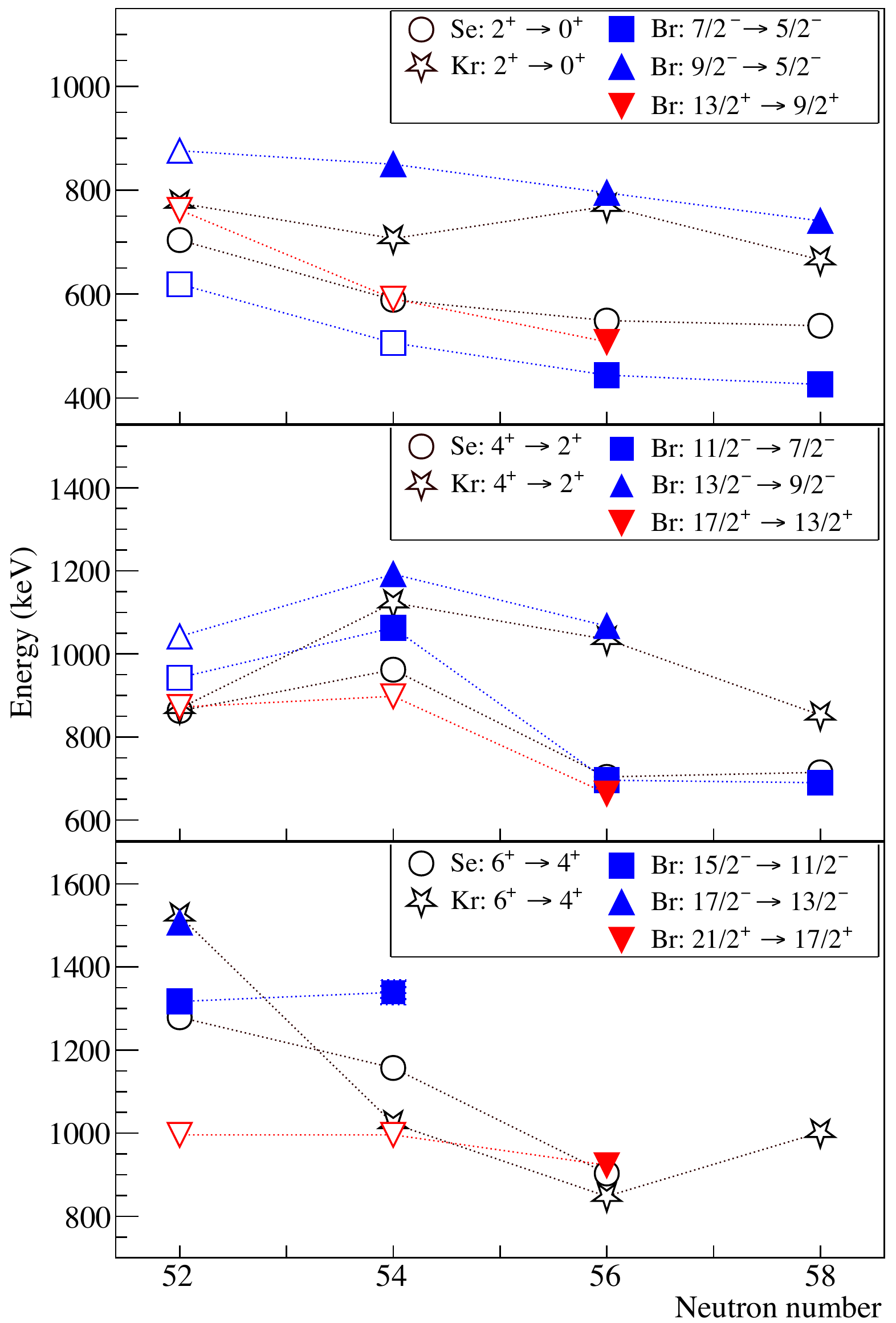}
    \caption{(color online) Comparison of the energies between Se, Kr and Br isotopes: the $2^+$ states in Se and Kr are compared to transitions $7/2^-\rightarrow5/2^-$, $9/2^-\rightarrow5/2^-$, and $13/2^+\rightarrow9/2^+$ in Br. The $4^+$ states are compared to $11/2^-\rightarrow7/2^-$, $13/2^-\rightarrow9/2^-$, and $17/2^+\rightarrow13/2^+$. The $6^+$ states are compared to $15/2^-\rightarrow11/2^-$, $17/2^-\rightarrow13/2^-$, and $21/2^+\rightarrow17/2^+$. Empty symbols correspond to already published measurements, filled symbols correspond to newly reported states in this work. Energy values are taken from~\cite{ensdf}.}
    \label{fig:system_Se}
\end{figure}

A comparison of the $2^+$, $4^+$ and $6^+$ energies of the even-even Se and Kr isotopes with Br excited states resulting from the discussed single particle to core couplings, for neutron numbers from 50 to 58, is shown in Fig.~\ref{fig:system_Se}. Empty symbols are taken from the literature~\cite{ensdf} while filled symbols are used for the states measured for the first time in this work. The energies of the positive parity states of the Br isotopes (shown in red downward-pointing triangles), are in overall good agreement with the Se systematics, especially for the $2^+$ and $4^+$ states, suggesting a very pure particle-core excitation of the $\pi g_{9/2}$ orbital, at low angular momentum. For the negative parity states, two different behaviors are observed. The transitions belonging to the band based on the $7/2^-$ state ($7/2^-\rightarrow5/2^-$, $11/2^-\rightarrow7/2^-$ and $15/2^-\rightarrow11/2^-$ -in blue squares), are also following the Se systematics, while the transitions belonging to the band based on the $9/2^-$ state ($9/2^-\rightarrow5/2^-$, $13/2^-\rightarrow9/2^-$ and $17/2^-\rightarrow13/2^-$ -in blue upward-pointing triangles) show a trend that is closer to the Kr systematics. It can be noted that while the $2^+$ state energy, and the one of the corresponding Br states, are slowly decreasing with the increasing neutron number, a sudden ``drop'' is observed for the $4^+$ and $6^+$ states at $N=56$.

\subsection{Constraints on the excited levels lifetimes}

As reported in detail in \cite{Astier_2006}, in the considered mass region, the $9/2^+$ state, coming from the promotion of the odd proton to the $\pi g_{9/2}$ sub-shell, is usually isomeric, because of the hindrance of its $\gamma$ decay. Indeed, the experimental work reported in~\cite{Simpson2010} reveals that such a state is long-lived in the $^{91,95}$Rb isotopes (16 and 94~ns, respectively). While disproved in~\cite{Astier_2006}, the possibility of a lifetime of about 20\,ns for the first excited $9/2^+$ state in \brseven has been suggested in the beta-decay study of~\cite{wisniewski_2019}, on the basis of $\gamma$ intensity balance evaluations. A search for isomeric states in neutron-rich odd-mass Br isotopes has thus been carried out in our work. In addition to providing information on excitation energies, both the FIPPS and the AGATA-VAMOS++ data-sets can be used to extract information regarding the lifetime of the excited states. On the one hand, the prompt-delayed coincidence analysis of the FIPPS data-set allows the identification of short-lived isomers and an estimation of their lifetime~\cite{Kandzia_2020, Iskra2020_PRC}. Such an analysis has been performed on \brseven revealing a non isomeric character of this state. On the other hand, the characteristics of the AGATA-VAMOS++ setup can be exploited to set an upper limit of $\sim 5$~ns to the lifetime of the observed states. Indeed, in such inverse kinematics fission reaction, after $\sim 5$~ns, the nucleus is ``outside'' the ``field of view'' of the AGATA array. Based on these considerations, it was concluded that none of the observed excited states had a lifetime $>5$~ns. This result is of particular importance for the $9/2^+$ state, as previously introduced.

\section{Theoretical calculations}\label{sec:theory}

\begin{figure}[htb] 
  \centering
  \includegraphics[width=0.75\linewidth]{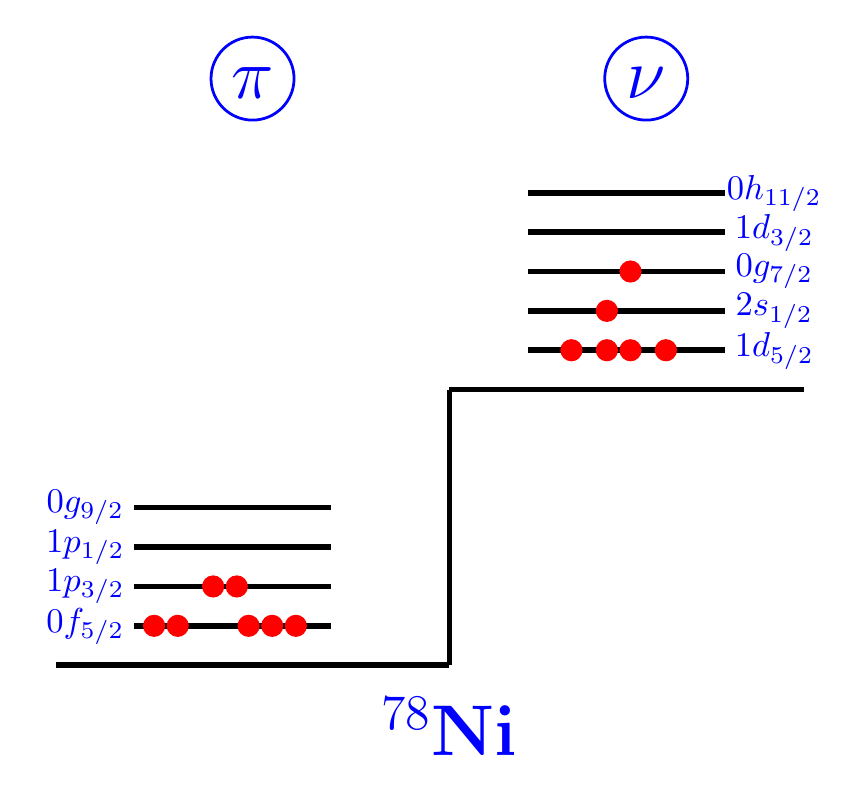}
  \caption{(color online) Valence space used in the present theoretical description, with an example of possible configuration of states of the basis, illustrated for the $^{91}$Br case.}
  \label{fig:Valence_space}
\end{figure}

In order to interpret the present experimental data, both Large Scale Shell Model (LSSM) and DNO-SM~\cite{Dao_2022} calculations have been performed. A similar theoretical study was recently presented in~\cite{Rezynkina_2022}, to describe the transition from semi-magicity to $\gamma$-softness in the $^{83-87}$As odd nuclei.  The shell-model valence space is spanned by the full $Z=28-50$ proton major shell and the full $N=50-82$ neutron major shell beyond $^{78}\text{Ni}$, namely, the  ($0f_{5/2}, 1p_{3/2}, 1p_{1/2}, 0g_{9/2}$) proton orbitals and the ($0g_{7/2}, 1d_{5/2}, 1d_{3/2}, 2s_{1/2}, 0h_{11/2}$) neutron orbitals. The valence space is illustrated in Fig.~\ref{fig:Valence_space} for the case of $^{91}$Br. The same effective interaction (DF2882) used in~\cite{Rezynkina_2022} with minor adjustments (see this latter reference  for more details) is employed here. It is a realistic two-body effective interaction with constraints on the monopole components, introduced to incorporate 3N forces effects. 

\begin{figure*}[htb]
    \centering
    \includegraphics[width=\linewidth]{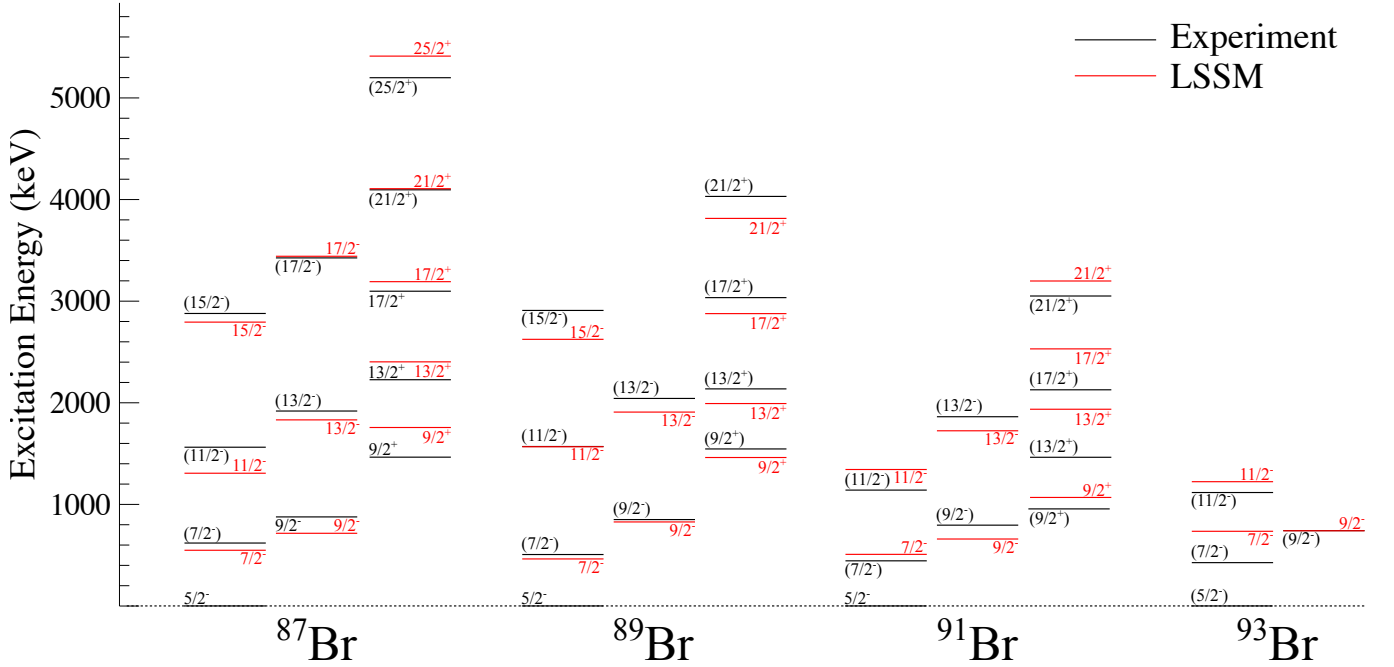}
   \caption{(color online) Global comparison between LSSM calculations and experiment for the yrast states of Br isotopes.}
    \label{fig:SM_calcs_comparison_yrast}
\end{figure*}

\begin{figure*}[htb]
    \centering
    \includegraphics[width=\linewidth]{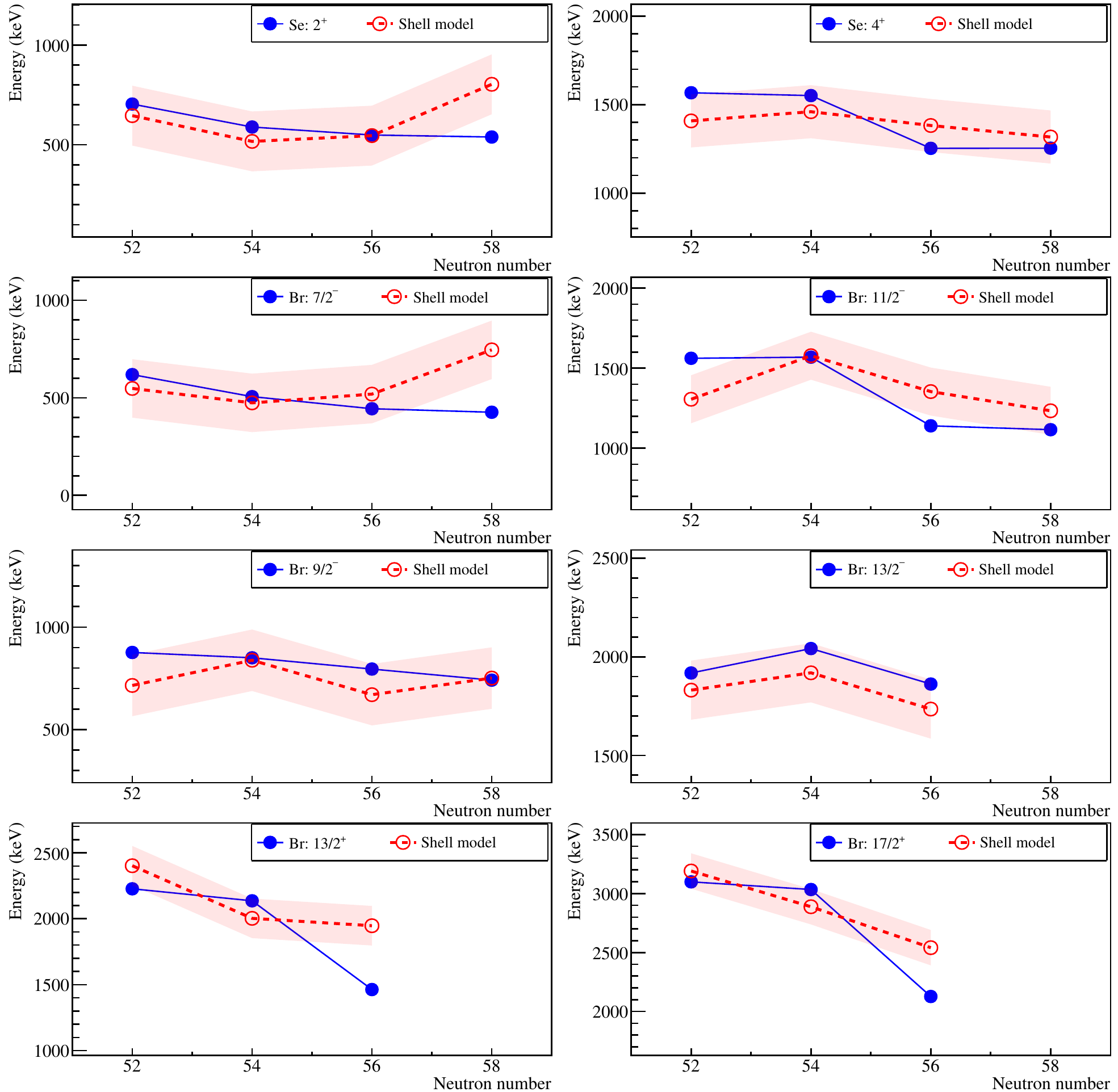}
   \caption{(color online) Comparison between LSSM calculations and experiment for a selection of levels in Se and Br isotopes. The energy intervals of $\pm 150$\,keV represented in the different pictures demonstrate the good agreement between calculations and experimental data.}
    \label{fig:SM_calcs_comparison}
\end{figure*}

A first comparison between LSSM calculations and the experimental data for all the yrast states measured in this work is shown in Fig.~\ref{fig:SM_calcs_comparison_yrast}, showing an overall very good agreement. A more detailed analysis of this comparison, for a selection of states for each band as a function of the neutron number, is done in Fig.~\ref{fig:SM_calcs_comparison}. The values for the $2^+$ and $4^+$ states in Se are also reported in this case. A band of $\pm 150$~keV is drawn on the theoretical values to guide the eye, figuring the theoretical error bars. A trend can be observed for the $2^+$ in Se isotopes, and the corresponding $7/2^-$ state in Br, showing that the LSSM calculations are ``overestimating'' the excitation energies at $N=58$, most likely due to the lack of solid information about the p-n residual interaction in this region. Moreover, the agreement is sensibly less good for the $13/2^+$ and $17/2^+$ state in \brnineonec. These specific cases apart, the overall agreement between calculations and experimental data is remarkably good. An additional comparison between the presented LSSM calculations and the experimental data can be seen in Fig.~\ref{fig:systemBr_9_2_p} where the predictions for the $9/2^+$ state energy are reported for the Br isotopic chain. A very good agreement is again observed and, in particular, the previously discussed energy ``drop'' at $N=56$ is reproduced. The calculations predict a very fragmented wave function, with only half of the $\nu d_{5/2}$ orbitals filled and the three remaining neutrons occupying the rest of the valence space.

\begin{figure}[htb]
    \centering
    \includegraphics[width=\linewidth]{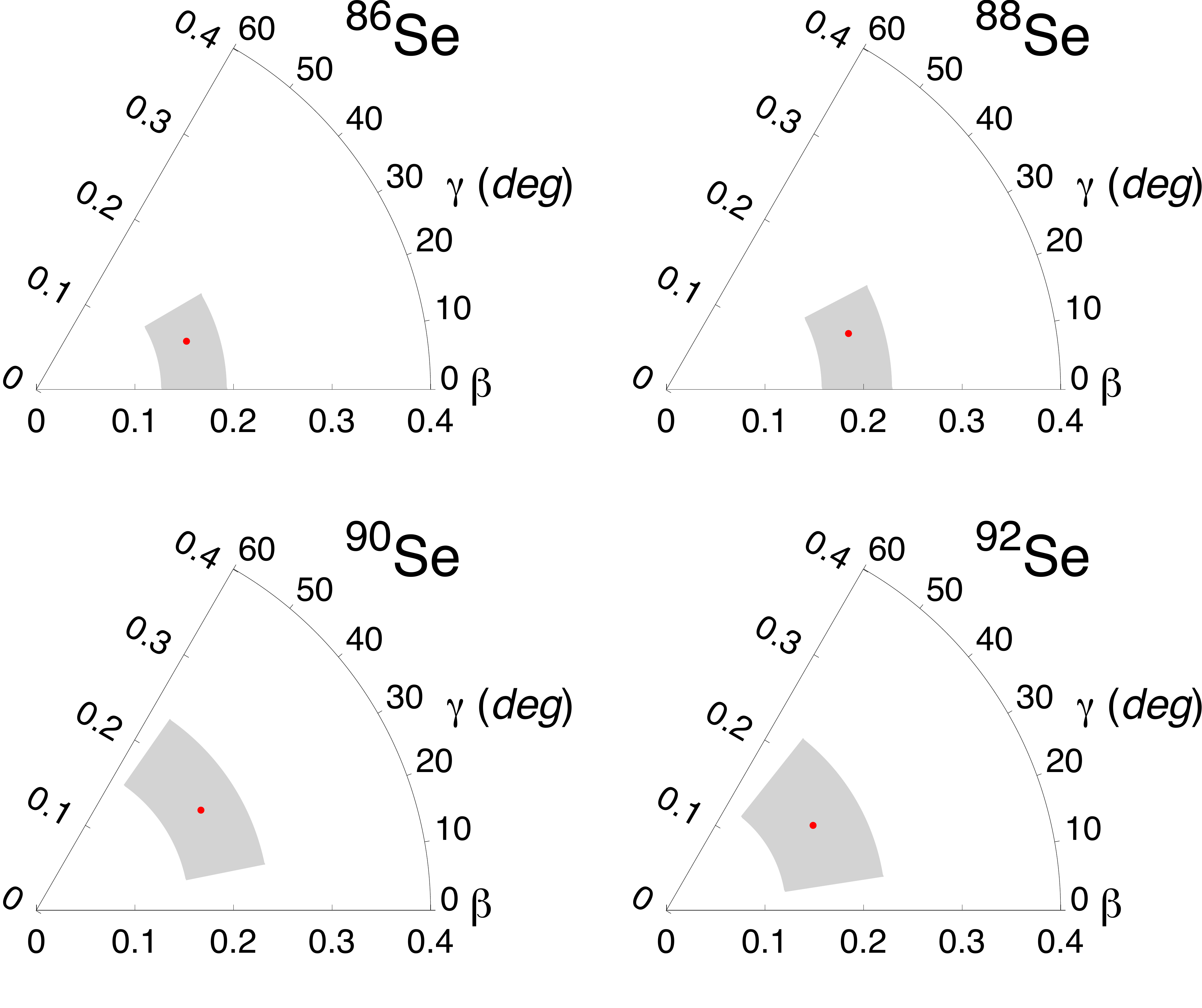}
    \caption{(color online) Kumar invariants for the $^{86, 88, 90, 92}$Se isotopes. Red dots represent the effective $\beta$, $\gamma$ values and grey shaded areas represent the 1$\sigma$ fluctuations contours.}
    \label{fig:kumar}
\end{figure}

In order to infer the degree of deformation of such isotopes from a Shell Model perspective, the Kumar invariants~\cite{Poves-PhysRevC.101.054307} for the neighbouring even-even Se isotopes can be investigated. Those are reported in Fig.~\ref{fig:kumar} for the $^{86, 88, 90, 92}$Se isotopes. This systematics demonstrates a transition from prolate shapes for $^{86,88}$Se to an onset of triaxiality for the $^{90, 92}$Se nuclei, with indeed very large fluctuations for these latter cases.

To further investigate the collectivity of those isotopes, the LSSM spectroscopic results can be compared and analysed in the framework of the Discrete Non-Orthogonal Shell Model (DNO-SM) recently developed in Ref.~\cite{Dao_2022}. The model uses the same valence space and allows diagonalization of the same Hamiltonian using the effective interaction DF2882 in a relevant deformed Hartree-Fock (HF) states basis from the potential energy surface represented in a $(\beta,\gamma)$ plane. The diagonalization is then performed after the rotational symmetry restoration using angular momentum projection technique. 

\begin{figure}[htb]
\centering\includegraphics[width=\linewidth]{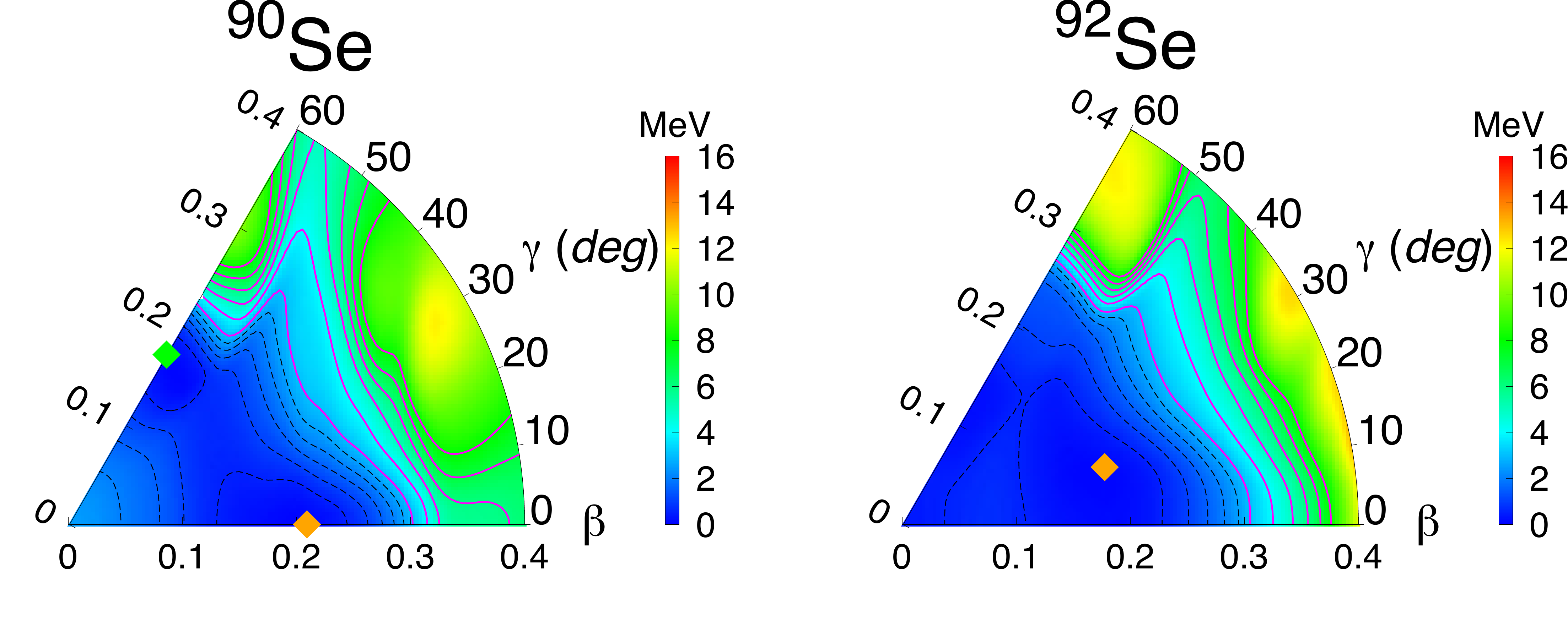}
\caption{Potential Energy Surface (PES) for the $^{90}$Se and $^{92}$Se isotopes. Orange and green diamonds indicate the location of prolate and oblate minima, respectively, when occurring.}
\label{fig:PES_Se}
\end{figure}

The potential energy surfaces (PES) obtained for $^{90}$Se and $^{92}$Se using the DNO approach are shown in Fig.~\ref{fig:PES_Se}. Such calculations suggest a very ``flat'' surface for both cases, with no dominant leading minima, allowing for competing prolate, oblate and triaxial regimes.

\begin{figure}[htb]
\centering\includegraphics[width=\linewidth]{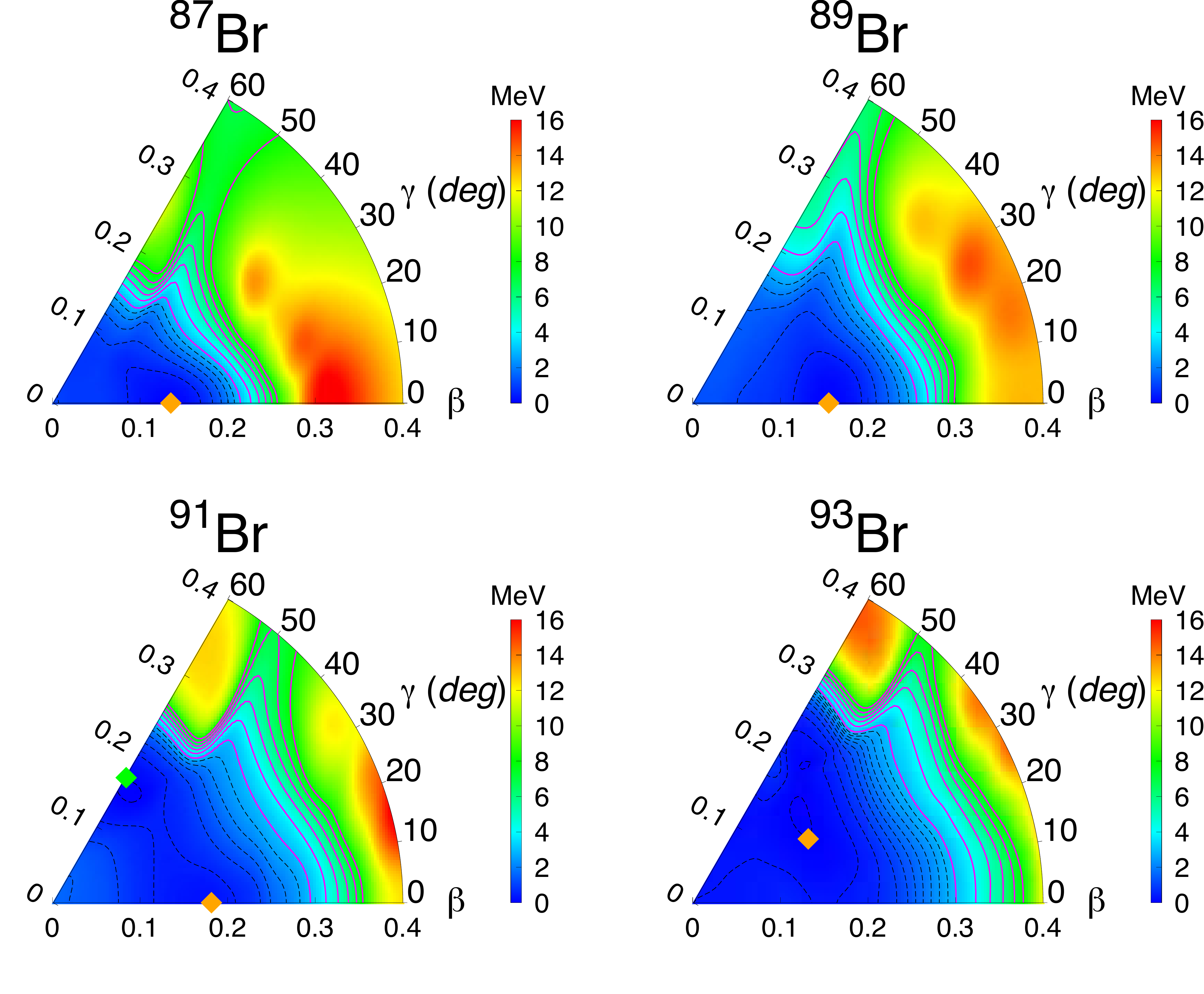}
\caption{Potential Energy Surface (PES) for the $^{87, 89, 91, 93}$Br. Orange and green diamonds indicate the location of prolate and oblate minima, respectively, when occurring.}
\label{fig:PES_Br}
\end{figure}

Fig.~\ref{fig:PES_Br} presents the PES of $^{87, 89, 91, 93}$Br. The onset of deformed configurations is observed in all cases. The first two cases show a moderate prolate HF minimum extending towards non-axial shapes. In the case of $^{91}$Br, two axial degenerated minima are found on the oblate and prolate axes,
while in $^{93}$Br, no shape is clearly favored with competing prolate, oblate and triaxial shapes like observed in the daughter $^{92}$Se case. 

\begin{figure*}[htb]
\centering\includegraphics[width=\linewidth]{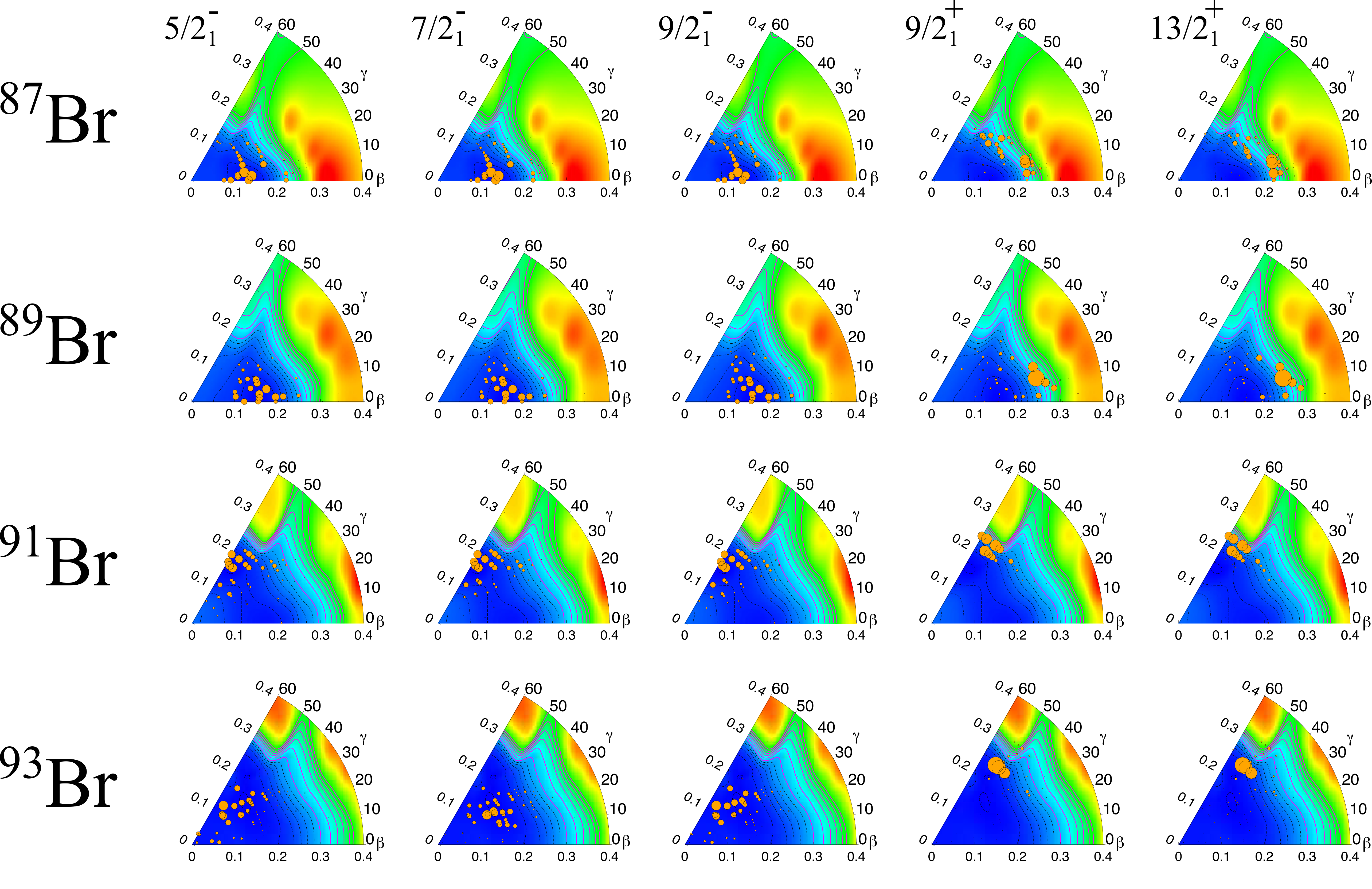}
\caption{(color online) Deformation structures of the yrast states in the $(\beta,\gamma)$ plane for $^{87, 89, 91, 93}$Br isotopes. The area of orange circles is directly proportional to the normalized probability to find a deformation $(\beta,\gamma)$ in the corresponding state.}
\label{fig:DNO_def}
\end{figure*}

All the considered states are then optimized through the mixing of deformed projected HF states, with the energy-minimization technique over the whole PES, starting from the corresponding HF minimum. The results of these calculations for $^{87}$Br, $^{89}$Br, $^{91}$Br and $^{93}$Br are shown in Fig.~\ref{fig:DNO_def} for the selected states $5/2^-_1$, $7/2^-_1$, $9/2^-_1$, $9/2^+_1$ and $13/2^+_1$. When moving from $^{87}$Br to $^{93}$Br, a clear transition from prolate to oblate deformation is observed. For $^{89}$Br, the negative parity states are quite mixed but always well localised in the vicinity of the prolate axis with $\beta$ values around $0.15$. The positive parity states are much purer, and well localised around the prolate axis with larger deformation ($\beta \sim 0.25)$. For $^{91}$Br, the negative and positive parity states are both localised on the oblate side of the $(\beta,\gamma)$ plane, with slightly larger deformation values ($\sim 0.18$ for the negative parity states and $\sim 0.25$ for positive parity states).  Finally in $^{93}$Br, as already inferred from the PES similarity with $^{91}$Br, one observes the same kind of deformation regime, with minor non-axial admixtures. 

\begin{figure}[htb] 
\hskip 0pt
    \includegraphics[width=1\linewidth]{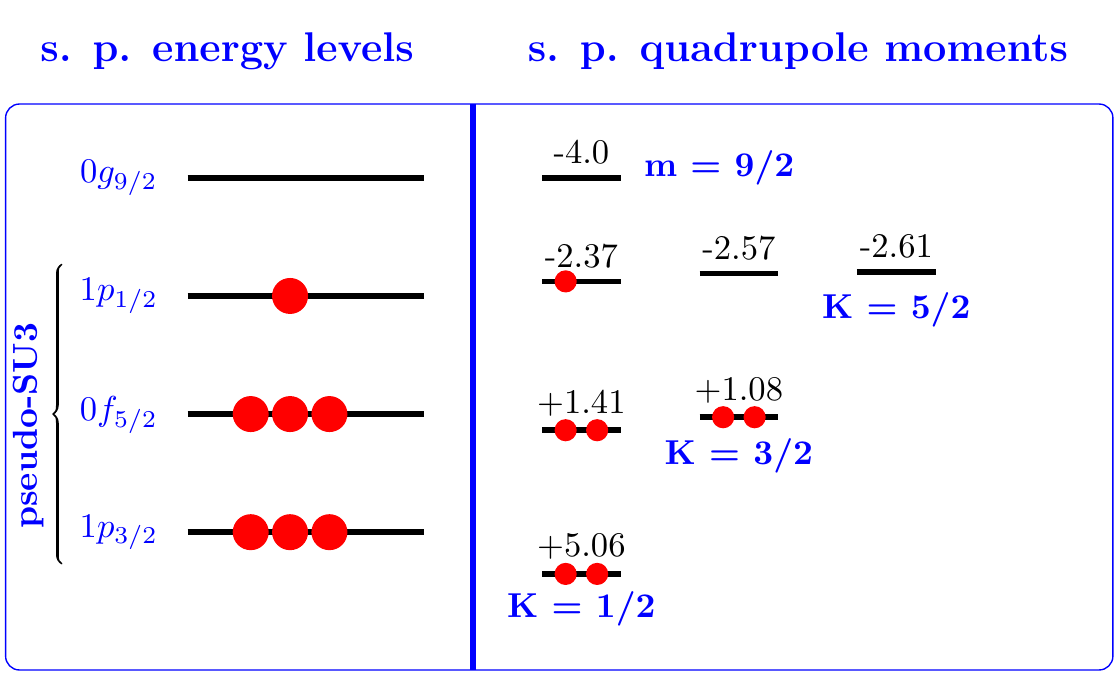}
    \includegraphics[width=1\linewidth]{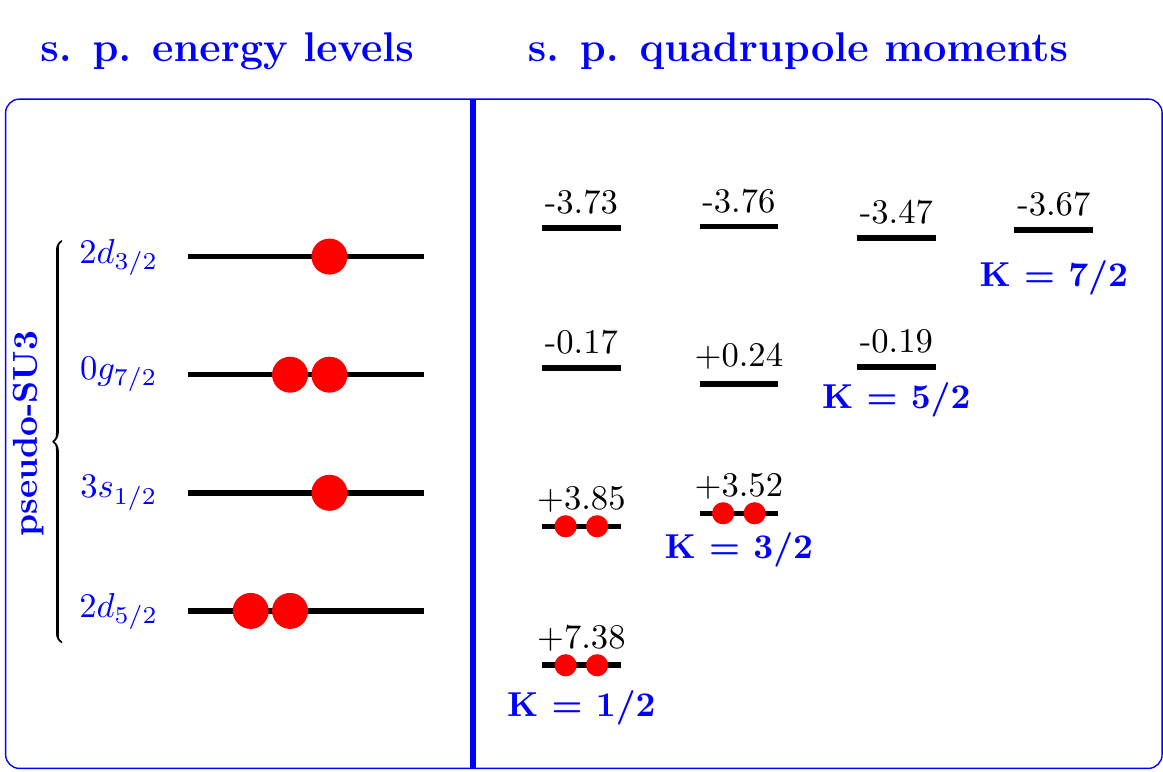}
  \caption{(color online) Zuker-Retamosa-Poves diagrams for the pseudo-SU3 proton orbitals (top) and neutron orbitals (bottom) limits in the case of $^{91}$Br prolate configuration. Positive quadrupole moment values correspond to prolate shape.}
  \label{fig:ZRP}
\end{figure}

In order to understand the above-described shape transition, collective behaviours can be inferred from the analysis of the quadrupole properties, on the basis of pseudo-SU3 symmetries, as shown in the recent arsenic investigation~\cite{Rezynkina_2022}. This implies that the $f_{5/2}, p_{3/2}$ and $p_{1/2}$ proton and $d_{3/2}, d_{5/2}, g_{7/2}, s_{1/2}$ neutron orbitals, with degenerate single-particle energies and a pure quadrupole-quadrupole interaction, are adopted. Following the procedure reported in~\cite{Zuker-PhysRevC.92.024320}, and using the Zuker-Retamosa-Poves (ZRP) diagrams (see Fig.~\ref{fig:ZRP} and Ref.~\cite{Zuker-PhysRevC.92.024320,NOWACKI2021103866}), the evolution of quadrupole correlations as a function of proton and neutron numbers can be estimated and configurations and quadrupole moments for the corresponding Br isotopes associated. 

It is worth recalling that in such a scheme, filling the Nilsson-SU3 proton orbitals produces maximum collectivity at mid-shell for six protons (half of the pseudo-SU3 space) where prolate and oblate configurations exhibit the same quadrupole moment of opposite sign ($\pm$ 15.10 e\,b$^2$), filling the shell either from the top or from the bottom. In Br isotopes, the additional proton  favors oblate configurations for both natural negative parity (-14.02 versus +12.73 e\,b$^2$), and non natural positive parity, adding the -4 e\,b$^2$ oblate quadrupole moment of a $g_{9/2}$ proton (see formula (6) and Tab.~II in~\cite{Zuker-PhysRevC.92.024320}), resulting in a value of -19.1 instead of +12.73 e\,b$^2$. On the neutron side, the filling of the pseudo-$pf$ SU3 Nilsson orbitals clearly favors prolate configurations for 2, 4 and 6 particles (+14.76, +22.46 and +29.5 e\,b$^2$), but it produces similar prolate and oblate quadrupole moments for 8 particles (+29.98 versus -29.26 e\,b$^2$). The predicted total quadrupole moments are summarized in Tab.~\ref{tab:QuadMom}. Taking into account the total effective charge ($e_p + e_n =2.4$), these values are in full agreement with all the DNO-SM calculations mentioned above, which exhaust up to 70\% of the SU3 limit in all cases.

\begin{table}[t]
\begin{ruledtabular}
\begin{tabular}{l rrrr}
& \multicolumn{4}{c}{$Q_0$ ( e\,b$^2$)}  \\[1pt]
                    & $\pi$         & $\nu$     & tot       & tot(eff)  \\
\hline \\ [-1pt]
\textbf{$^{87}$Br}  &               &           &           &           \\
prolate             &  +12.73       &  +14.76   & +27.49    & +66.98    \\ 
oblate              &   -14.02      &  -7.52    & -21.54    & -51.70    \\
oblate + $g_{9/2}$  &   -19.10      &  -7.52    & -25.54    & -61.30    \\
\textbf{$^{89}$Br}  &               &           &           &           \\
prolate             &  +12.73       & +22.46    & +35.19    & +84.46    \\
oblate              &   -14.02      &  -14.98   & -29.00    & -69.60    \\ 
oblate + $g_{9/2}$  &   -19.10      &  -14.98   & -33.00    & -79.20    \\ 
\textbf{$^{91}$Br}  &               &           &           &           \\
prolate             & +12.73        &  +29.50   & +42.23    & +101.35   \\
oblate              &  -14.02       &  -22.32   & -36.34    & -87.22    \\ 
oblate + $g_{9/2}$  &  -19.10       &  -22.32   & -41.42    & -99.41    \\
\textbf{$^{93}$Br}  &               &           &           &           \\
prolate             & +12.73        &  +29.98   & +42.71    &  +102.50  \\
oblate              &  -14.02       &  -29.26   & -43.28    &  -103.87  \\ 
oblate + $g_{9/2}$  &  -19.10       &  -29.26   & -48.36    & -116.06   \\
\end{tabular}
\end{ruledtabular}
\caption{Nilsson-SU3 intrinsic quadrupole moments $Q_0$ estimates (in units of e\,b$^2$) for the odd $^{87-93}$Br . Positive and negative $Q_0$ are referred to prolate and oblate shapes, respectively. The total effective values correspond to multiplying bare proton and neutron quadrupole moments by a $(e_p+e_n=)2.4$ factor.}
\label{tab:QuadMom} 
\end{table}

The combination of these results for the proton and neutron values exhibits two features: a rise of quadrupole collectivity with increasing neutron number but also a shape transition from prolate shape in $^{87, 89}$Br to prolate/oblate shape coexistence in $^{91}$Br and oblate shape in $^{93}$Br, as seen in Fig.~\ref{fig:PES_Br}. As evident from Tab.~\ref{tab:QuadMom}, the SU3 limits are never realized and they are affected by the shell evolution and levels splitting. As a consequence, in the case of flat PES landscape for $^{91,93}$Br, the pseudo-scheme may be altered to favor oblate shapes over the prolate ones, as inferred from the DNO-SM analysis.

A similar behavior has already been predicted at $N=58$ in neutron-rich Kr isotopes, on the basis of Symmetry Conserving Configuration Mixing (SCCM) calculations with the Gogny D1S interaction~\cite{Rodriguez2014}.

In summary, the overall theoretical description points to clear deformation in rotational regime for $^{87}$Br, $^{89}$Br, $^{91}$Br and $^{93}$Br. The maximum of deformation is achieved in $^{91}$Br ($N=56$), in agreement with the increasing deformation observed in the Selenium isotopes by Chen \textit{et al.}~\cite{Chen_2017} (from  $^{84}$Se to  $^{90}$Se). It is worth emphasising that, while the arsenic isotopes in the region show a $\gamma$-soft regime~\cite{Rezynkina_2022}, the present study points to a shape transition from prolate $^{87, 89}$Br to oblate $^{91, 93}$Br for both natural and non-natural parity bands. 

\section{Conclusion}\label{sec:conclusion}
New spectroscopic information for neutron-rich Br isotopes has been obtained from fission experiments performed at two different $\gamma$-ray spectroscopy setups. The isotopic selectivity of the VAMOS++ spectrometer, combined with the Doppler correction capabilities of the AGATA array, have been exploited to obtain new spectroscopic information up to the exotic \brninethreec. The statistics available at FIPPS, using neutron-induced fission and an active target, has allowed for the confirmation of existing literature in \brseven and \brninec, as well as for the placement of new transitions in \brninec. A new upper limit of 5~ns has been obtained from the analysis of the two data-sets, on the lifetime of the $9/2^{+}$ state in the considered isotopes. 

The obtained experimental results have been discussed by analysing the systematics in the mass region, in particular the one of the even-even Se isotopes. Evidences for particle-core excitations have been highlighted and the considered systematics has also been used for spin-parity assignments of newly found excited states. 

Through advanced Shell-Model and DNO Shell-Model calculations, evidences for a shape transition from prolate $^{87, 89}$Br to oblate $^{91, 93}$Br for both natural and non-natural parity bands are shown. This evolution is intimately related to the delicate competition between proton and neutron quadrupole excitations, exhibiting a ``showcase'' of Nilsson-SU3 algebraic model~\cite{Zuker-PhysRevC.92.024320}. As the region of isotopes with $26\leq Z \leq 38$ and $42\leq N \leq 60$ has shown to be decisive for first-peak r-process abundances in neutron-star merger calculations~\cite{Reiter-PhysRevC.101.025803}, this calls for extended experimental and theoretical studies of such neutron-rich isotopes. The direct measurement of quadrupole moments, for example, would be extremely beneficial. 

\section*{Acknowledgments}

We acknowledge the AGATA and FIPPS collaborations for the setup and data collection of the experimental campaigns, as well as the important technical contributions of G.~Fremont, J.~Goupil, B.~Jacquot, L.~Ménager, J.~Ropert, C.~Spitaels, and the GANIL accelerator staff. The authors are grateful to A.O.~Machiavelli for the fruitful discussions. Work partially funded by MICIU  MCIN/AEI/10.13039/501100011033, Spain with grants PID2020-118265GB-C42, PRTR-C17.I01, Generalitat Valenciana, Spain with grant CIPROM/2022/54, ASFAE/2022/031,  CIAPOS/2021/114 and by the EU NextGenerationEU, ESF funds. I.~K. and D.~S. were supported by the National Research, Development and Innovation Fund of Hungary (NKFIH), financed by the project with contract No. TKP2021-NKTA-42, as well as under the K18 funding scheme with project No. K147010. C.~Mihai and collaborators from IFIN-HH acknowledge Nucleu project No. PN 23 21 01 02. 

\clearpage

\bibliography{MainFile}

\end{document}